\documentclass[useAMS,usenatbib]{mn2e}
\usepackage{epsfig}

\title[Long-term monitoring in IC4665]{Long-term monitoring in IC4665: 
Fast rotation and weak variability in very low mass objects}

\author[Scholz, Eisl{\"o}ffel \& Mundt]{Alexander Scholz$^{1}$\thanks{E-mail:
as110@st-andrews.ac.uk}, Jochen Eisl{\"o}ffel$^{2}$\thanks{E-mail: jochen@tls-tautenburg.de} 
and Reinhard Mundt$^{3}$\\
$^{1}$SUPA, School of Physics \& Astronomy, University of St. Andrews, North Haugh, St. Andrews, 
Fife KY16 9SS, United Kingdom\\
$^{2}$Th{\"u}ringer Landessternwarte Tautenburg, Sternwarte 5, D-07778 Tautenburg, Germany\\
$^{3}$Max-Planck-Institut f{\"u}r Astronomie, K{\"o}nigstuhl 17, D-69117 Heidelberg, Germany}

\begin{document}

\date{Accepted. Received.}

\pagerange{\pageref{firstpage}--\pageref{lastpage}} \pubyear{2002}

\maketitle

\label{firstpage}

\begin{abstract}
We present the combined results of three photometric monitoring campaigns targeting very low mass (VLM) stars
and brown dwarfs in the young open cluster IC4665 (age $\sim 40$\,Myr). Each of our observing runs covers timescales 
of $\sim 5$ days in the seasons 1999, 2001, 2002, respectively. In all three runs, we observe 
$\sim 100$ cluster members, allowing us for the first time to put limits on the evolution of spots and 
magnetic activity in fully convective objects on timescales of a few years. For 20 objects covering masses from 
0.05 to 0.5$\,M_{\odot}$ we detect a periodic flux modulation, indicating the presence of magnetic spots co-rotating 
with the objects. The detection rate of photometric periods ($\sim 20$\%) is significantly lower than in solar-mass 
stars at the same age, which points to a mass dependence in the spot properties. With two exceptions, none of the 
objects exhibit variability and thus spot activity in more than one season. This is contrary to what is seen in 
solar-mass stars and indicates that spot configurations capable of producing photometric modulations occur 
relatively rarely and are transient in VLM objects. The rotation periods derived in this
paper range from 3 to 30\,h, arguing for a lack of slow rotators among VLM objects. The periods fit into a 
rotational evolution scenario with pre-main sequence contraction and moderate (40-50\%) angular momentum 
losses due to wind braking. By combining our findings with literature results, we identify two regimes of 
rotational and magnetic properties, called C- and I-sequence. Main properties on the C-sequence are fast 
rotation, weak wind braking, H$\alpha$ emission, and saturated activity levels, while the I-sequence is 
characterised by slow rotation, strong wind braking, no H$\alpha$ emission, and linear activity-rotation 
relationship. Rotation rate and stellar mass are the primary parameters that determine in which regime an 
object is found. We outline a general scheme to understand rotational evolution for low-mass objects in 
the context of these two regimes and discuss the potential as well as the problems of this scheme.
\end{abstract}


\section{Introduction}

Rotation and magnetic activity are key parameters of stellar evolution. While these 
parameters have been well-explored for solar-mass stars over the past decades, our knowledge 
is still sparse in the very low mass (VLM) regime. The interest in rotation and activity 
in VLM stars and brown dwarfs is fueled by the hypothesis that these properties depend 
on a fundamental level on the interior structure. Specifically, to operate the type of dynamo 
that generates the solar-type cyclical magnetic field, the presence of a shear layer 
between convective envelope and radiative core is thought to be of prime importance
\citep[e.g.][and references therein]{1980Natur.287..616S,1982A&A...106...58S}. 
Objects with masses $\la 0.3\,M_{\odot}$, however, are fully convective throughout their 
evolution \citep{1997A&A...327.1039C}. Thus, if the aforementioned assumptions are correct, 
they would not be able to generate magnetic fields in the same way as solar-type stars. Changes 
in magnetic properties and, consequently, rotational evolution are therefore expected at the 
fully-convective boundary \citep[see][]{1998A&A...331..581D}. 

Rotation and activity are interdependent in a twofold sense: On one side, rotation plays 
an important role in driving the dynamo and essentially powers magnetic activity. On the 
other side, stellar winds driven by magnetic fields are the predominant agent of the spindown 
on the main-sequence and thus determine the long-term rotational evolution of stars. 
For solar-mass stars, the close connection between rotation and activity manifests itself in the 
rotation/activity relation, which is linear for slow rotators and flattens at high rotation 
rates, as well as in the Skumanich law \citep{1972ApJ...171..565S}, which establishes that 
rotation rates and activity indicators both decline with the square root of age on the 
main-sequence.

Over the past decade, it has been shown convincingly that, with the exception of a few very young 
objects, VLM stars and brown dwarfs are fast rotators with periods generally shorter than 2\,d \citep[e.g.][]{2001A&A...367..218B,2004A&A...421..259S,2007prpl.conf..297H,2008ApJ...684.1390R,2009AIPC.1094..118R}.
The periods found in the Monitor project confirm this finding \citep[see in particular][]{2008MNRAS.383.1588I}.
Rotational braking due to stellar winds is clearly less efficient in this mass regime. On the
other hand, a number of ultracool objects are now known to harbour strong, large-scale 
magnetic fields, as evidenced by Zeeman Doppler imaging \citep{2008MNRAS.390..545D,2008MNRAS.390..567M}, 
direct field measurements \citep{2007ApJ...656.1121R}, and radio observations 
\citep[e.g.][]{2006ApJ...648..629B,2006ApJ...653..690H}. A full theory that accounts for all 
these findings is still missing.

Investigating rotation and activity in the VLM regime therefore has the potential to probe 
fundamental physical questions: To what extent does interior structure affect the observable 
properties of stars? How does magnetic field generation change as a function of object mass? 
Can we successfully predict rotation rates for a given age and mass? The latter problem leads
to the intriguing possibility to use rotation periods -- a quantity that can be measured with
high accuracy -- to determine stellar ages and masses \citep[gyrochronology, see][]{2007ApJ...669.1167B}.

The goal of this paper is to provide new observational constraints on rotation and activity in
the very low mass regime. We will probe a specific subset of magnetic phenomena, namely the
properties of magnetically induced spots in the photosphere. We present results from three photometric
monitoring campaigns in the young open cluster IC4665, including 20 new rotation periods for VLM stars 
and brown dwarfs (Sect. \ref{monitor}-\ref{tsa}). Through repeated monitoring, we are able to probe the 
characteristics and the long-term evolution of magnetic spots in this critical mass regime 
(Sect. \ref{spots}). Finally, we will use our periods in combination with literature data to probe 
rotational braking and magnetic properties in low-mass objects (Sect. \ref{rotation} and \ref{probs}).

\section{Photometric monitoring and data reduction}
\label{monitor}

\subsection{Observations}
\label{obs}

The target field for this variability study covers parts of the central region of the
open cluster IC4665. The field is centred at $17^h45^m 15.0^s \delta +05^o 21'17\farcs0$ 
(J2000.0) and is covered by our photometric survey (Eisl{\"o}ffel et al., in prep., see Sect. 
\ref{targets}). This region was observed in three I-band monitoring campaigns covering a time span 
of four years in total, using the ESO/MPG 2.2-m Wide Field Imager at La Silla and the 2-m Schmidt telescope 
at the Th{\"u}ringer Landessternwarte (TLS). Both instruments have a comparable field of view of 
$\sim 0.32$\,sqdeg. 

\begin{table}
    \caption[]{Observing log for the three observing runs, runs A and C were carried out
    with the ESO/MPG 2.2-m WFI, run B with the TLS Schmidt camera. Columns 3-6 contain
    number of exposures, integration time per exposure, sky conditions, median seeing.}
       \label{ts} 
       \begin{tabular}{llcclc}
	  \hline
          run & date  & no. & exp. time & weather & seeing\\
          \hline
          A & 28/05/1999 & 1  & 600\,s & photometric   & 1\farcs2 \\
            & 05/06/1999  & 44 & 500\,s & some cirrus   & 0\farcs7 \\
	    & 06/06/1999  & 43 & 500\,s & photometric   & 0\farcs9 \\
	    & 08/06/1999  & 3  & 500\,s & partly cloudy & 1\farcs2 \\
	    & 09/06/1999  & 5  & 500\,s & photometric   & 1\farcs2 \\
	  \hline
	  B & 19/05/2001 & 13 & 600\,s & partly cloudy & 3\farcs4 \\
	    & 20/05/2001  & 24 & 600\,s & some cirrus   & 2\farcs7 \\
	    & 21/05/2001  & 23 & 600\,s & partly cloudy & 2\farcs9 \\
	    & 22/05/2001  & 23 & 600\,s & photometric   & 2\farcs6 \\
	    & 23/05/2001  & 22 & 600\,s & photometric   & 2\farcs6 \\
	  \hline
	  C & 29/05/2002 & 15 & 500\,s & photometric   & 1\farcs6 \\
	    & 07/06/2002  & 14 & 500\,s & partly cloudy & 1\farcs2 \\ 
	    & 08/06/2002  & 14 & 500\,s & photometric   & 1\farcs3 \\
	    & 09/06/2002  & 10 & 500\,s & some cirrus   & 1\farcs5 \\
	    & 12/06/2002 & 17 & 500\,s & some cirrus   & 1\farcs6 \\
	    & 13/06/2002 & 17 & 500\,s & some cirrus   & 1\farcs3 \\
       \end{tabular}
\end{table}

Details for the three observing runs are given in Table \ref{ts}. In the first WFI run in 
May/June 1999, we took 96 long exposure images of this field, including complete coverage 
of two consecutive nights. The second WFI run was carried out in service mode, resulting 
in dense coverage of typically 2-3\,h per night for six nights distributed over two weeks.
This campaign provided 87 time series images of our field in IC4665. 
The TLS observing run covers five consecutive nights with typically 4\,h of continuous 
observations per night, in total 105 exposures. The TLS time series suffered from 
bright sky background and sub-optimal seeing, and thus the photometric accuracy 
was significantly lower than in the WFI runs. The typical integration time for individual
exposures was 500\,sec in all three runs.

\begin{figure*}
\includegraphics[width=4.0cm,angle=-90]{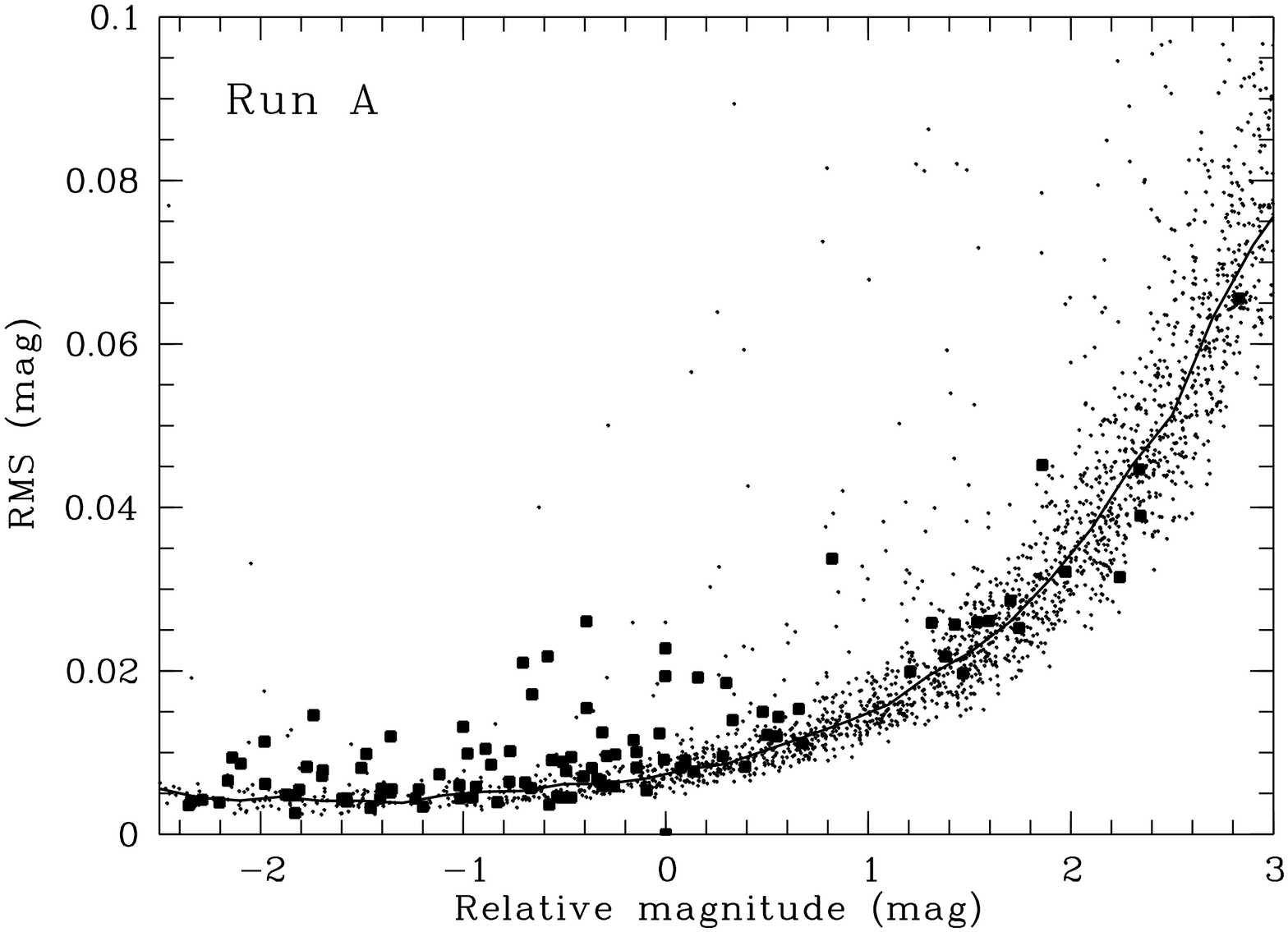} \hfill
\includegraphics[width=4.0cm,angle=-90]{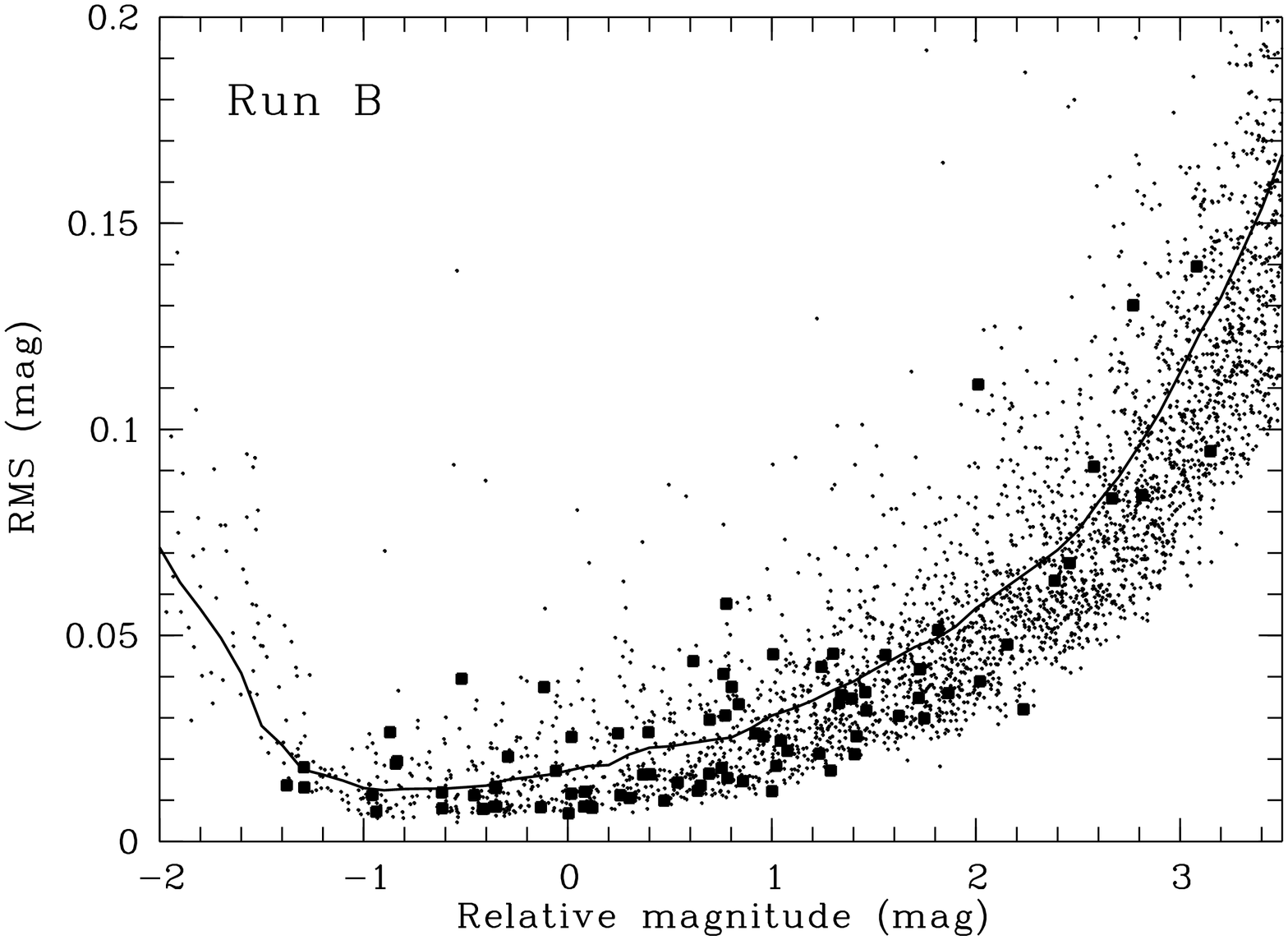} \hfill
\includegraphics[width=4.0cm,angle=-90]{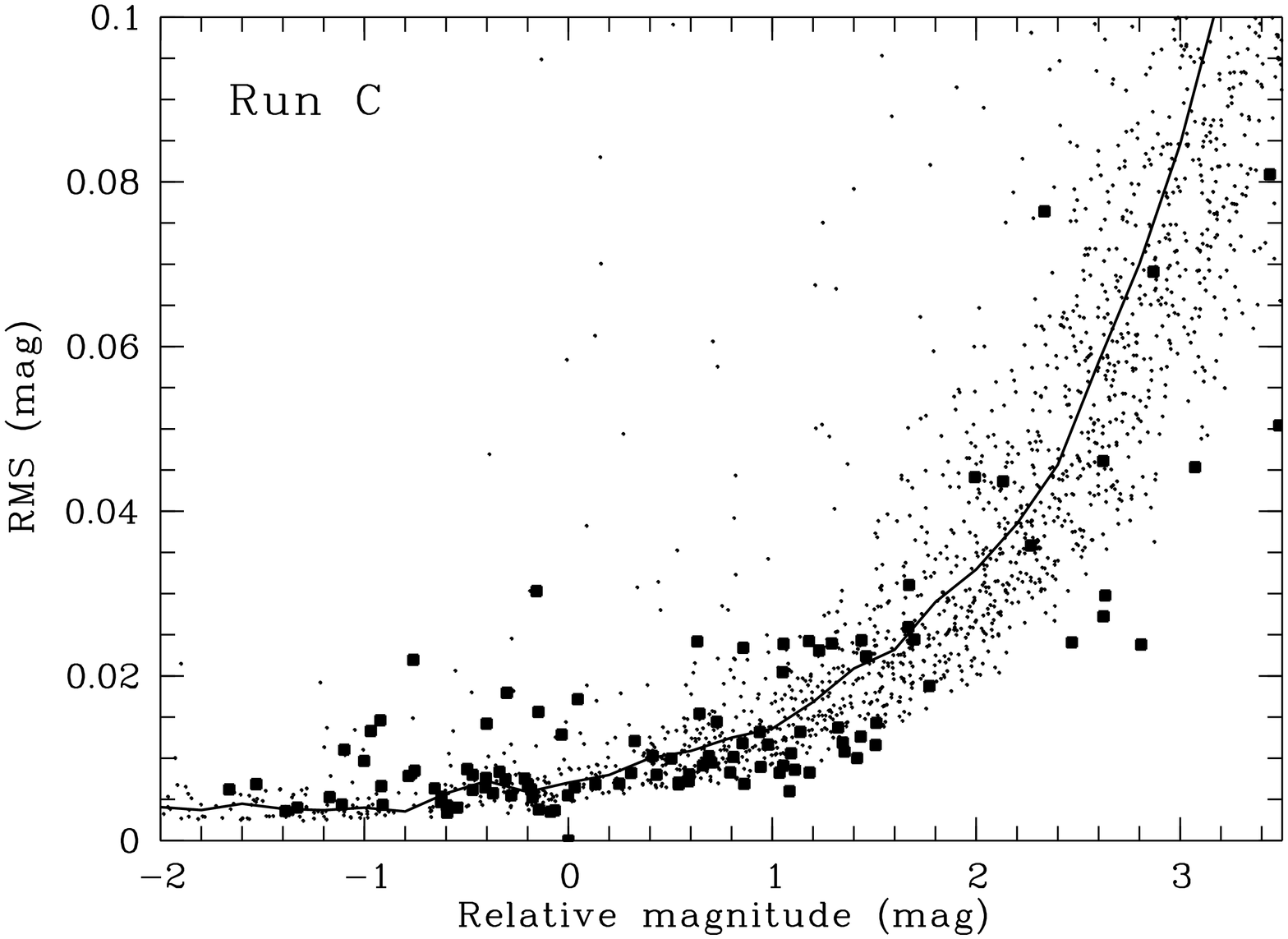} 
\caption{RMS vs. relative magnitude for all three runs: solid lines show the binned median rms, large symbols
the VLM candidate members. Note the different scale on the y-axis of the middle panel. For run B, we only plot 
4000 of 10000 field objects, for clarity. Due to the differences in the I-band filters used in the various campaigns, 
the relative magnitudes are not easily transformed to calibrated I-band photometry. The approximate I-band 
mags for the zeropoint on the relative scale are 17.2, 16.2, and 16.5\,mag for runs A-C, respectively,
with an uncertainty of $\pm 0.3$\,mag. These numbers are derived by comparing the relative magnitudes with
the absolute photometry derived in our survey (Sect. \ref{targets}).
\label{f1}}
\end{figure*}

\subsection{Image reduction}

All images were reduced in a similar fashion, following the recipes outlined in 
\citet{2004A&A...419..249S,2005A&A...429.1007S}. The reduction includes bias subtraction 
and flatfield correction using either twilight flats (WFI) or domeflats (TLS). Additionally, 
the I-band images have to be corrected for multiplicative large-scale illumination gradients, 
a consequence of imperfect flatfielding, and additive, small-scale fringe structures. 
Both effects can be corrected using a 'superflat' constructed from images of dark 
sky regions. By median-filtering the superflat, large-scale structures (illumination) 
can be separated from the small-scale fringes. As a result, we obtain an image of the
average illumiation and the average fringe pattern. Since 
fringes are variable as a function of airmass, temperature, and weather conditions, 
an appropriate scaling is required for each time series image. We used the automated 
scaling procedure outlined in \citet{2005A&A...429.1007S}, and manually refined the resulting 
scaling factors if necessary. We estimate the residuals after fringe correction to be not
larger than 1\% of the sky background.

\subsection{Instrumental magnitudes}

We derived instrumental magnitudes for all objects in the time series field
by fitting their point spread function (PSF) using the {\tt daophot} package
within IRAF \citep{1987PASP...99..191S}. For the WFI runs, all photometry steps were 
done on each chip of the mosaic separately to account for the variable properties 
of the CCDs, e.g. the nonlinearity of the chip response.

Model PSFs were calculated using typically 40 isolated stars. Then, this model was 
fit to each object in the field. The reliability of the resulting photometry was 
checked by subtracting all PSF fits from the original image. As result we obtained 
essentially empty frames, in which only saturated objects, bright galaxies, and some 
close undetected binary companions remained.

\subsection{Relative calibration}

The lightcurves of all objects were calibrated relatively with an average lightcurve
from a set of non-variable reference stars in the observed field. The selection of
reference stars was performed in a similar fashion for all three campaigns, using the
procedure described in \citet{2004A&A...419..249S}. In short, the routine compares each 
potential reference star with the average of all other potential reference stars. For the 
WFI runs, this procedure was again carried out for each chip of the mosaic separately. 
Typically, the final sample of reference stars comprises of $\sim 100$ objects per chip.
From these objects, the final reference lightcurve was determined. By subtracting this 
reference lightcurve from all time series from the respective observing run, we obtained 
relatively calibrated magnitudes for all objects.

In Fig. \ref{f1} a-c we plot the lightcurve RMS vs. relative magnitude for all three runs. 
Overplotted is the median rms per magnitude bin as solid line after excluding outliers 
-- this function gives the typical photometric error at a given relative magnitude. For 
bright, unsaturated stars with I-band magnitudes of 15-16, we obtain average uncertainties 
of 5-8\,mmag in runs A and C and 10-20\,mmag in run B. Overplotted in large symbols in Fig. 
\ref{f1} are the datapoints for the sample of cluster member candidates, described in Sect. 
\ref{targets}. 

\section{The sample of probable cluster members}
\label{targets}

\begin{figure}
\includegraphics[width=6cm,angle=-90]{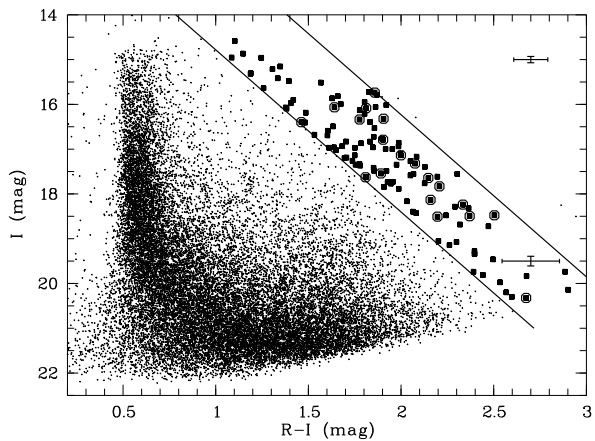} 
\caption{I,R-I colour magnitude diagram for our time series field in IC4665: The two solid lines 
illustrate the colour criterium applied to select the primary sample. All probable cluster members
according to their colours are shown as large dots; objects with periodic lightcurve are marked
with squares. Errorbars indicate typical photometric uncertainties. For clarity, only $\sim 10000$ 
out of $\sim 30000$ total datapoints are shown. Photometry obtained during run A (see Table \ref{ts}). 
Missing from that figure is object no. 119, which was identified in a different campaign (see text). 
\label{f2}}
\end{figure}

The targets for our variability analysis are 113 sources classified as probable
cluster members in our deep photometric survey in IC4665. The detailed outcomes of this survey 
will be published in a forthcoming paper (Eisl{\"o}ffel et al., in prep.). In short: We surveyed 
$\sim 0.4$ sqdeg of the central cluster in the R- and I-band using the ESO/MPG WFI at the 2.2-m 
telescope at La Silla -- these images have been obtained in the same observing run that provided 
the first time series (run A, see Sect. \ref{obs}). The R- and I-band data is calibrated
by comparing with photometry of Landolt standard fields observed in the same nights \citep{1992AJ....104..340L}.
A deep (I,R-I) colour magnitude diagram yielded 
a sample of 118 objects whose optical colours are consistent with an empirically determined 
isochrone for the cluster IC4665 (see Fig. \ref{f2}). We complemented the survey with near-infrared 
photometry, either from 2MASS or from our own J-band data\footnote{Observations carried out with the 
near-infrared camera $\Omega '$ at the 3.5-m telescope on Calar Alto/Spain, Deutsch-Spanisch 
Astronomisches Zentrum}. From the primary sample, 112 objects show near-infrared 
colours in agreement with the model isochrone for an age of 40\,Myr from \citet{1998A&A...337..403B}.

\begin{table*}
\caption{Basic data (coordinates, photometry, masses) for the cluster member candidates with 
periodic lightcurve in at least one of the monitoring campaigns. Full survey results to be 
published in Eisl{\"o}ffel et al. (in prep.). "2M" indices designate photometry from the 
2MASS database. Mass estimates derived by comparing the near-infrared photometry with 
evolutionary tracks by \citet{1998A&A...337..403B}.}
\label{cands}
\begin{tabular}{rlllllllll}
\hline
ID & $\alpha$ & $\delta$ & R & I & J & J$_{2M}$ & H$_{2M}$ & K$_{2M}$ & Mass\\
     & (J2000.0) & & (mag) & (mag) & (mag) & (mag) & (mag) & (mag) & ($M_{\odot}$)\\
\hline
  7& 17 43 59.64& 5 17 37.8&    19.43&    17.54&    15.90& 15.82 & 15.32 & 14.95 & 0.14\\
 28& 17 44 30.17& 5 18 49.2&    17.86&    16.40&    14.89& 14.91 & 14.33 & 13.93 & 0.27\\
 31& 17 44 31.46& 5 34 09.3&    20.70&    18.51&    16.68&       &       &       & 0.09\\
 36& 17 44 42.84& 5 19 24.8&    20.29&    18.13&    15.94& 16.05 & 15.28 & 15.25 & 0.12\\
 37& 17 44 43.23& 5 09 20.6&    19.79&    17.64&    15.67& 15.78 & 15.06 & 15.00 & 0.14\\
 39& 17 44 44.08& 5 21 09.7&    17.59&    15.74&    13.90& 13.93 & 13.31 & 13.00 & 0.51\\
 40& 17 44 46.45& 5 32 01.9&    20.04&    17.83&    16.16& 15.77 & 15.28 & 14.89 & 0.14\\ 
 49& 17 44 54.68& 5 23 52.4&    17.71&    16.07&    14.24& 14.49 & 13.78 & 13.56 & 0.37\\
 55& 17 45 05.57& 5 11 11.0&    18.11&    16.34&    14.67& 14.67 & 14.01 & 13.78 & 0.32\\
 57& 17 45 06.69& 5 30 38.4&    18.69&    16.79&    15.28& 15.07 & 14.41 & 14.01 & 0.25\\ 
 59& 17 45 10.95& 5 33 28.5&    19.14&    17.14&    15.53& 15.43 & 14.67 & 14.49 & 0.19\\ 
 63& 17 45 13.53& 5 18 39.2&    17.90&    16.08&    14.33& 14.32 & 13.73 & 13.36 & 0.41\\
 73& 17 45 21.39& 5 34 37.8&    19.39&    17.32&    15.73& 15.48 & 14.80 & 14.59 & 0.18\\
 75& 17 45 26.00& 5 29 40.5&    20.98&    18.48&    16.56& 16.21 & 15.63 & 15.24 & 0.11\\
 76& 17 45 27.20& 5 33 57.8&    23.00&    20.32&    18.00&       &       &       & 0.05\\ 
 88& 17 45 37.44& 5 23 40.5&    19.44&    17.63&    15.88& 15.95 & 15.27 & 15.00 & 0.13\\ 
100& 17 45 49.89& 5 09 40.2&    18.23&    16.32&    14.61& 14.66 & 13.97 & 13.64 & 0.34\\
110& 17 45 59.35& 5 29 31.6&    20.57&    18.24&    16.32& 16.01 & 15.44 & 15.12 & 0.12\\
118& 17 46 06.29& 5 36 39.3&    20.87&    18.50&         &       &       &       & 0.09\\
119& 17 45 37.10& 5 23 08.1&    15.79&         &         & 14.28 & 13.63 & 13.34 & 0.42\\  
\hline
\end{tabular}
\end{table*}

Furthermore, we surveyed a larger area of the cluster with the 2-m Schmidt telescope at the
Th{\"u}ringer Landessternwarte Tautenburg in a run in May 1999. The I-band photometry from this
campaign combined with 2MASS near-infrared data yielded an (I,I-J) colour-magnitude diagram. Applying
a colour selection, we obtained an additional sample of 57 probable cluster members. Only one of these 
sources is located in the monitoring field and shows a significant period in run A of the variability 
study (object \#119, see below). We include this object in the discussion. Thus, the sample 
for the time series analysis comprises 113 objects, with calibrated photometry in optical and near-infrared bands.

The photometrically selected cluster member candidates may be contaminated by
red field objects. The contamination rate was estimated based on optical colour-magnitude diagrams 
simulated using the Galaxy model presented by \citet{1986A&A...157...71R} and \citet{2003A&A...409..523R}. 
We derive a contamination rate of $\sim 30$\% for the total candidate sample, most of the contaminating
objects are expected to be M dwarfs in the fore- or background of the cluster. From the space density
for M dwarfs given by \citet{1997AJ....113.2246R} and our survey area follows a contamination rate of 
about 30\%, consistent with the Galaxy model.

%For these objects, the likelihood to be true cluster members may be
%higher than in the total candidate sample. This can be motivated based on the results 
%from \citet{2007AcA....57..149K}: They monitor 180 nearby M dwarfs and find periods for 31 of them, only 
%15 with periods $<4$\,d, which could have been detected in our monitoring campaigns (see Sect. \ref{perser}). 
%This is a 'success rate' of $\sim 8$\%. Assuming that 30\% of our total candidate sample are contamination, 
%we would expect to find $\sim 3$ of them to show periodic variations, corresponding to $\sim 15$\% of our 
%periodic sample. Based on this argument, the contamination among our objects with periodic variability would 
%be lower than in the total sample ($\sim 15$\% vs. 30\%). We note that the period sample by 
%\citet{2007AcA....57..149K} is dominated by early M dwarfs, whereas our targets are mostly expected 
%to have mid-M types, based on their colours. As shown by \citet{1998A&A...331..581D} from $v\sin i$
%data, the rotation rates in field dwarfs increase significantly at spectral type M3-M5. If this
%implies a higher detection rate for photometric periods than in early M dwarfs is not clear yet.

In Fig. \ref{f2} we show the (I,R-I) colour magnitude diagram, probably cluster members are marked. 
In our three monitoring campaigns, we found 20 objects with significant periodicity in the lightcurve 
(see Sect. \ref{tsa}); these objects are marked with open circles. As can be seen in this plot, the majority 
of the periodic objects forms a tight sequence with less scatter than the total sample, as expected for objects 
on the cluster isochrone. This may be seen as evidence for lower contamination among the objects with period,
although for a definite decision spectroscopy is required. 

In Table \ref{cands} we provide the basic data for all objects in our survey with period detection
in at least one of the monitoring campaigns. We estimate masses for these objects by comparing
their near-infrared magnitudes with the cluster isochrone from Baraffe et al. (1998), assuming
an age of 40\,Myr, a distance of 350\,pc, and an extinction of $E_{B-V}=0.18$ 
\citep{1981A&AS...44..467M,1981A&A....97..235M}.
According to this estimate, the objects span a mass range from 0.05 to 0.5$M_{\odot}$ and thus include
a few brown dwarf candidates. Due to uncertainties in the theoretical 
evolutionary tracks and differences between the standard bands and the filters used
in the survey, resulting in calibration offsets, the mass estimates may be inaccurate by as much
as 100\%, particularly at very low masses. Since both effects are systematic, the masses
for our objects are comparable on a relative scale. 

\section{Time series analysis}
\label{tsa}

The main focus of the time series analysis was the search for periodic variabilities.
The period search is described in detail in Sect. \ref{perser}. In addition, we 
looked for conspicuous signs for clear non-periodic variability in the lightcurves: 
All candidate lightcurves were inspected visually to register obvious signs of rapid 
variability. In particular, we searched for signs of flares, i.e. for brightness eruptions 
with gradual decline lasting several hours. No such events with $>0.1$\,mag amplitude 
were found, indicating that flare events are rare in VLM objects. Based on our 
time coverage, we estimate an I-band flare rate of $<3.2\cdot 10^{-4}\,h^{-1}$, in agreement 
with the result obtained in a previous paper 
\citep[$2.3\cdot 10^{-4}\,h^{-1}$,][]{2005A&A...429.1007S}.

In the RMS vs. magnitude plots in Fig. \ref{f1}, we mark the cluster members with larger 
symbols, to allow an assessment of the generic variability. As can be seen in these plots,
VLM objects in IC4665 show only low-level photometric variations, with amplitudes $<0.05$\,mag.
Most of the objects that are observed to have an RMS significantly larger than the field 
average are later identified as objects with periodic variability. This is in contrast to
the findings for field L dwarfs with similar masses, where variable objects are rarely
found to exhibit a photometric period \citep[e.g.,][]{2001A&A...367..218B,2002ApJ...577..433G}.

\subsection{Period search for runs A and C}
\label{perser}

We used the Scargle periodogram \citep{1982ApJ...263..835S} to identify candidate periods in the
lightcurves from runs A and C. For the initial detection, we relied on a preliminary significance 
estimate based on the equation given by \citet{1986ApJ...302..757H}. All candidate periods are then 
verified in a number of independent tests. In particular, periodograms and lightcurves of possible 
periodic objects were compared with those of nearby stars and excluded if the neighbours show a 
similar periodicity. In addition, we CLEANed the periodograms for all periodic objects from sampling 
artefacts and aliases using the procedure given by \citet{1987AJ.....93..968R} and checked if the peak 
found in the Scargle periodogram is still present. Only periods which pass all tests are accepted. 
The period search procedure is described in detail in \citet{2004A&A...419..249S} and has been 
thoroughly tested in previous campaigns \citep{2004A&A...421..259S,2005A&A...429.1007S}. 

\begin{figure*}
\includegraphics[width=3cm,angle=-90]{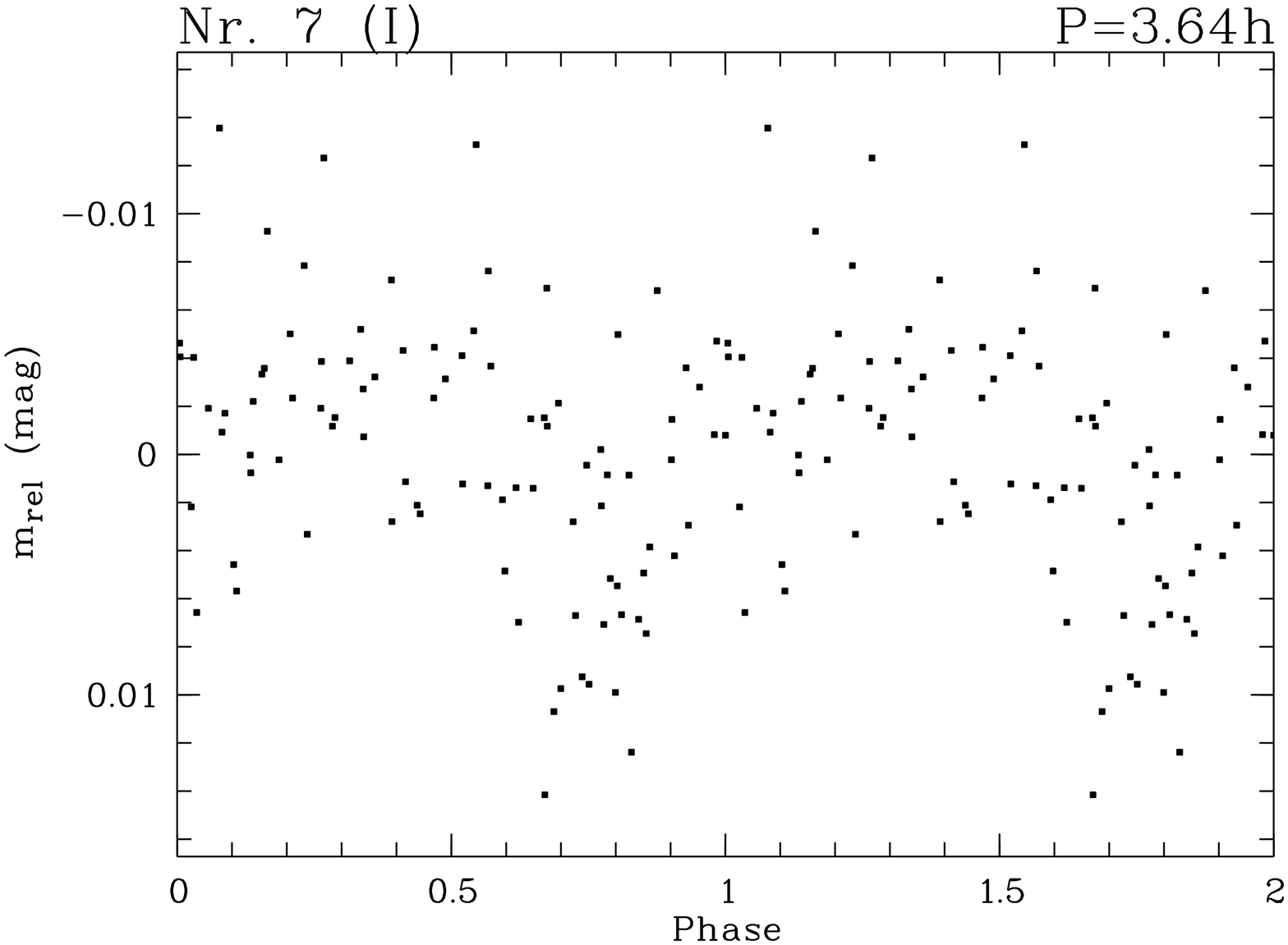} \hfill
\includegraphics[width=3cm,angle=-90]{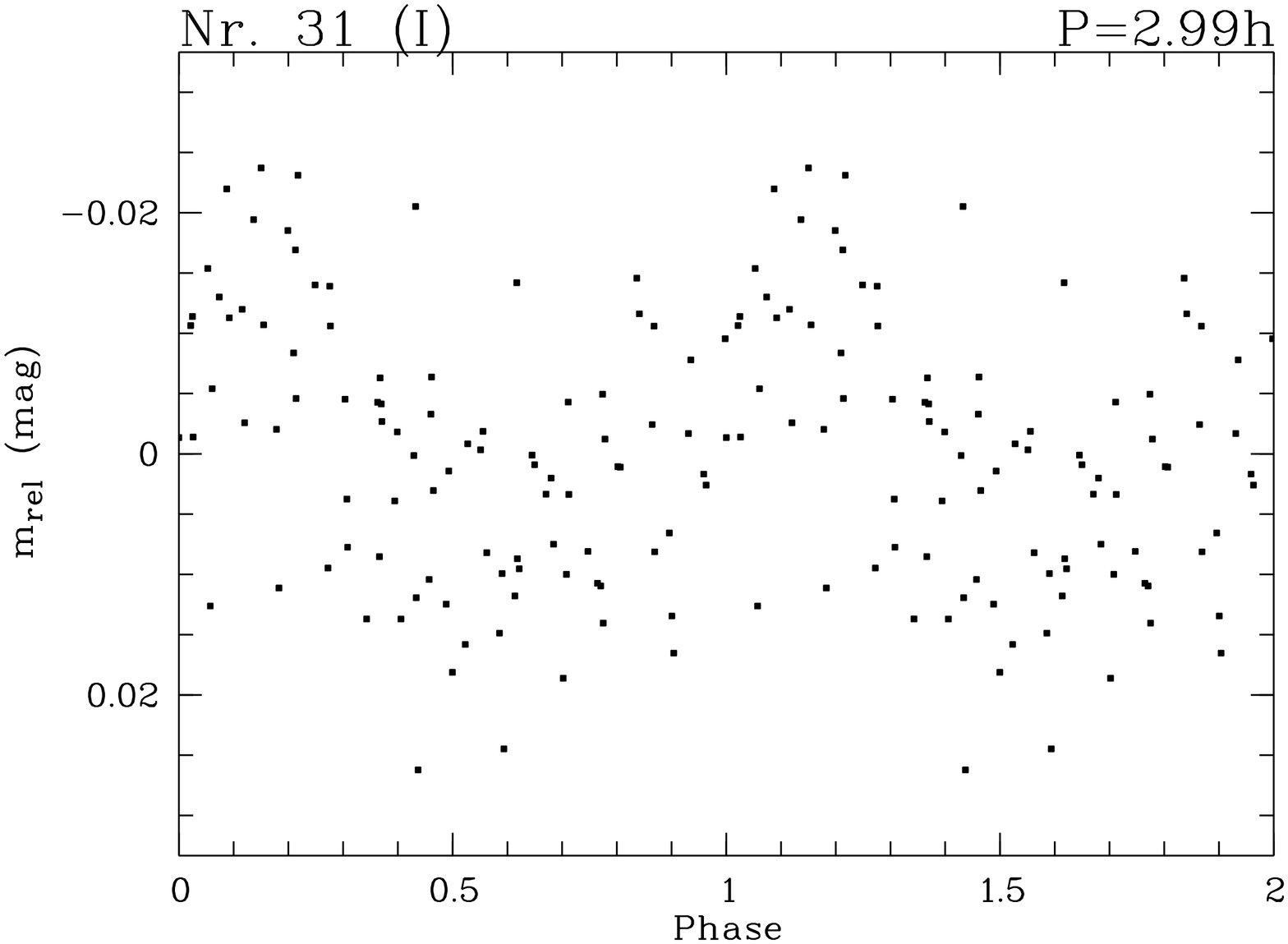} \hfill
\includegraphics[width=3cm,angle=-90]{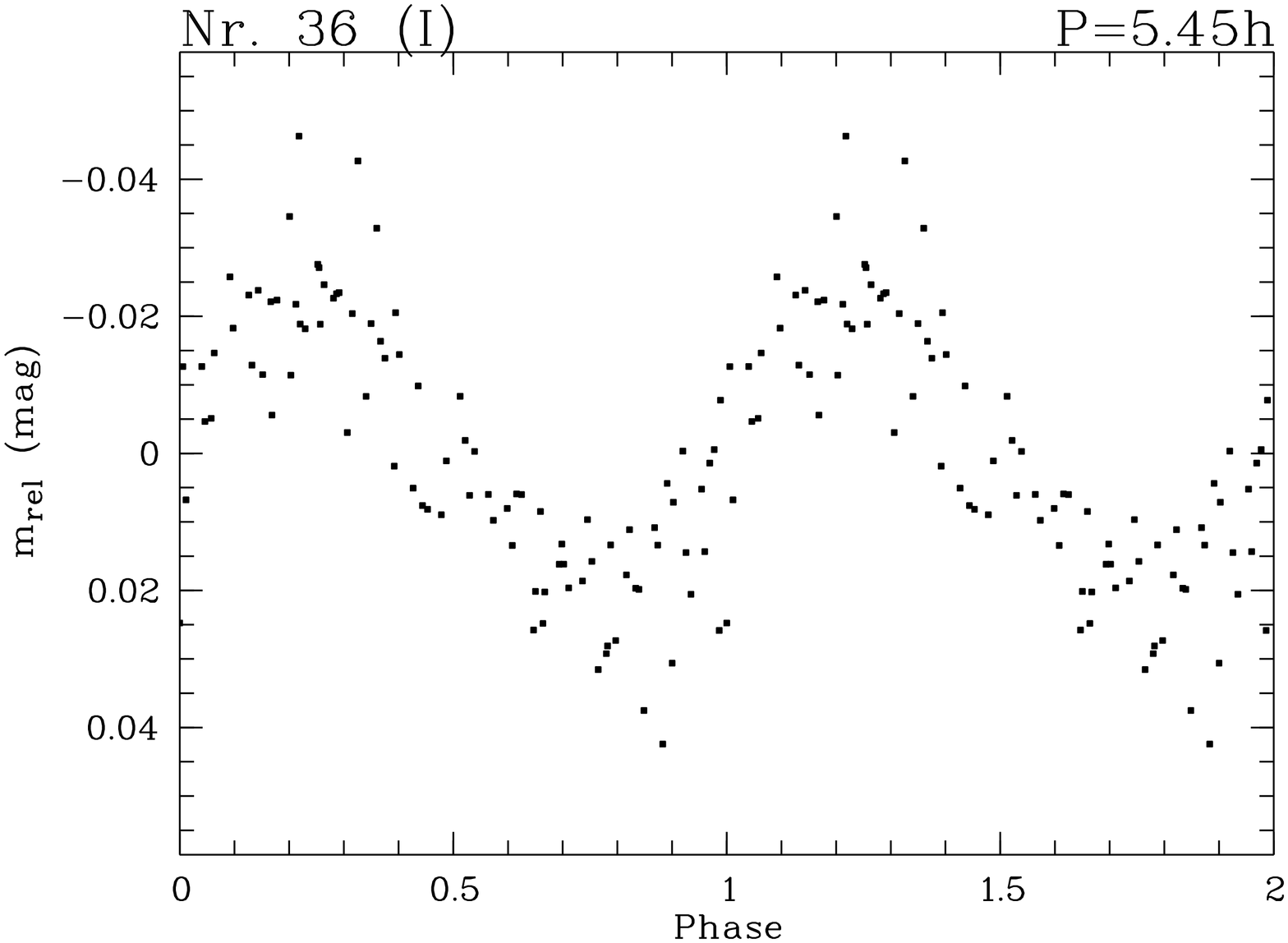} \hfill
\includegraphics[width=3cm,angle=-90]{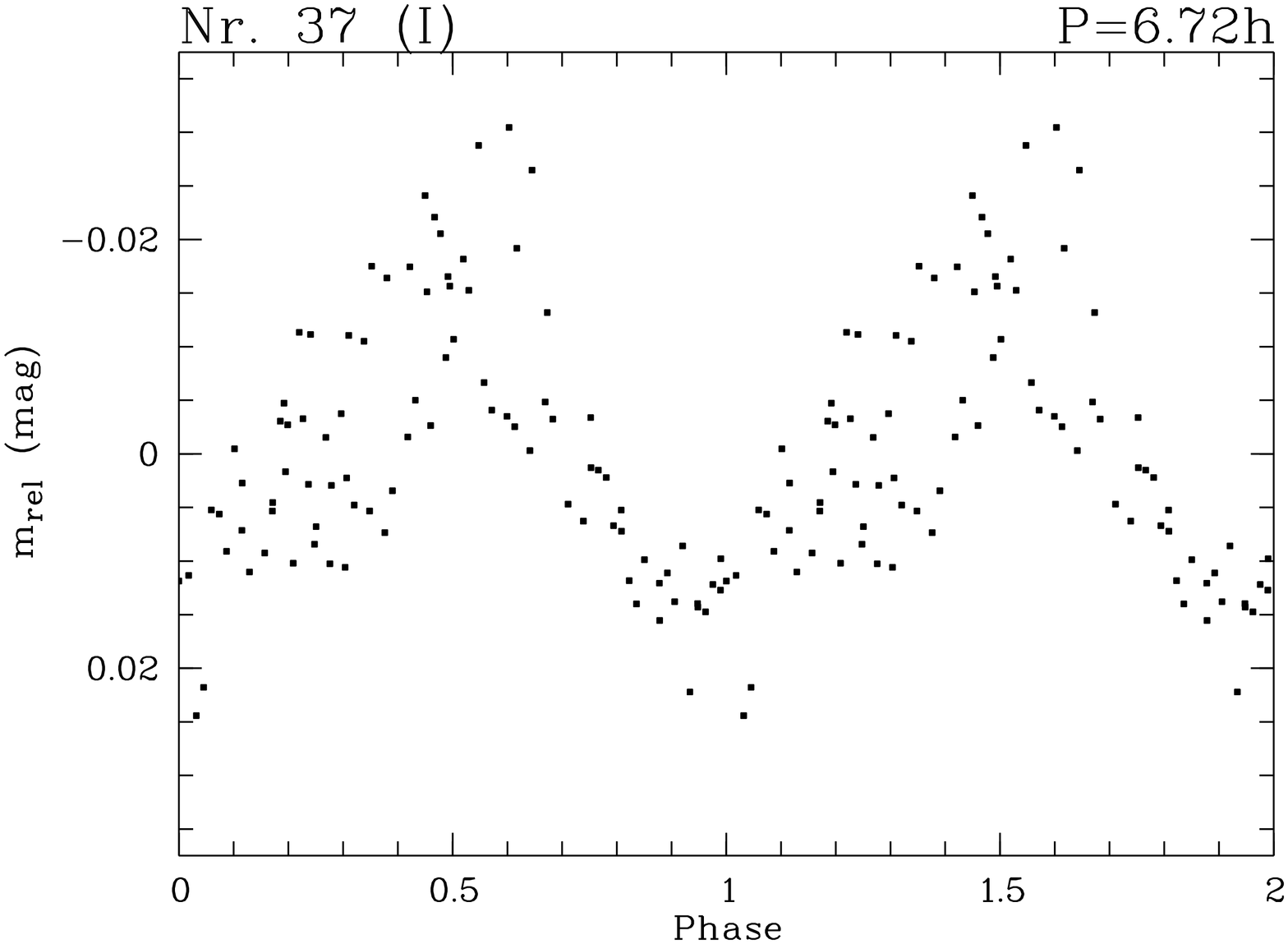} \\
\includegraphics[width=3cm,angle=-90]{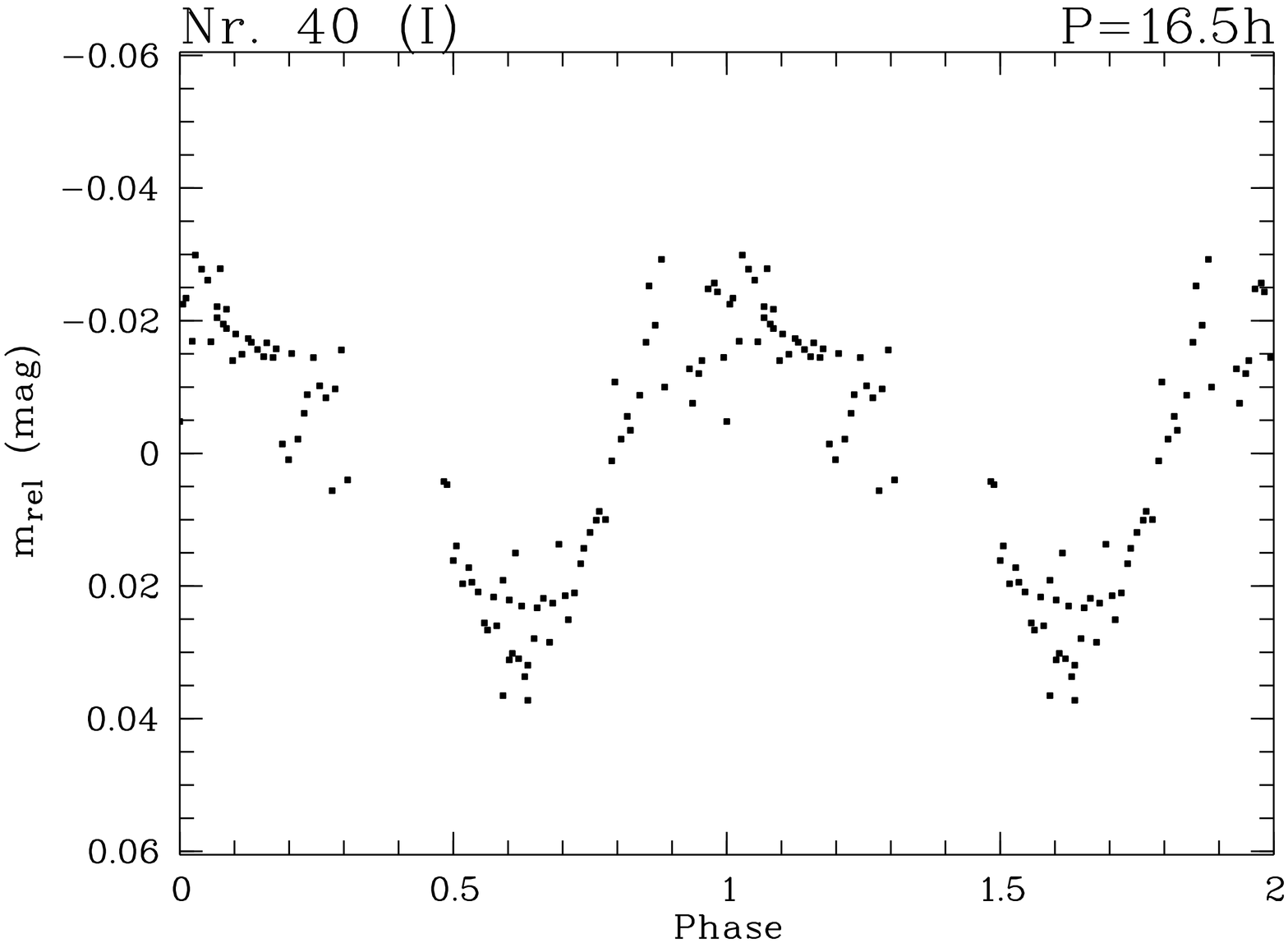} \hfill
\includegraphics[width=3cm,angle=-90]{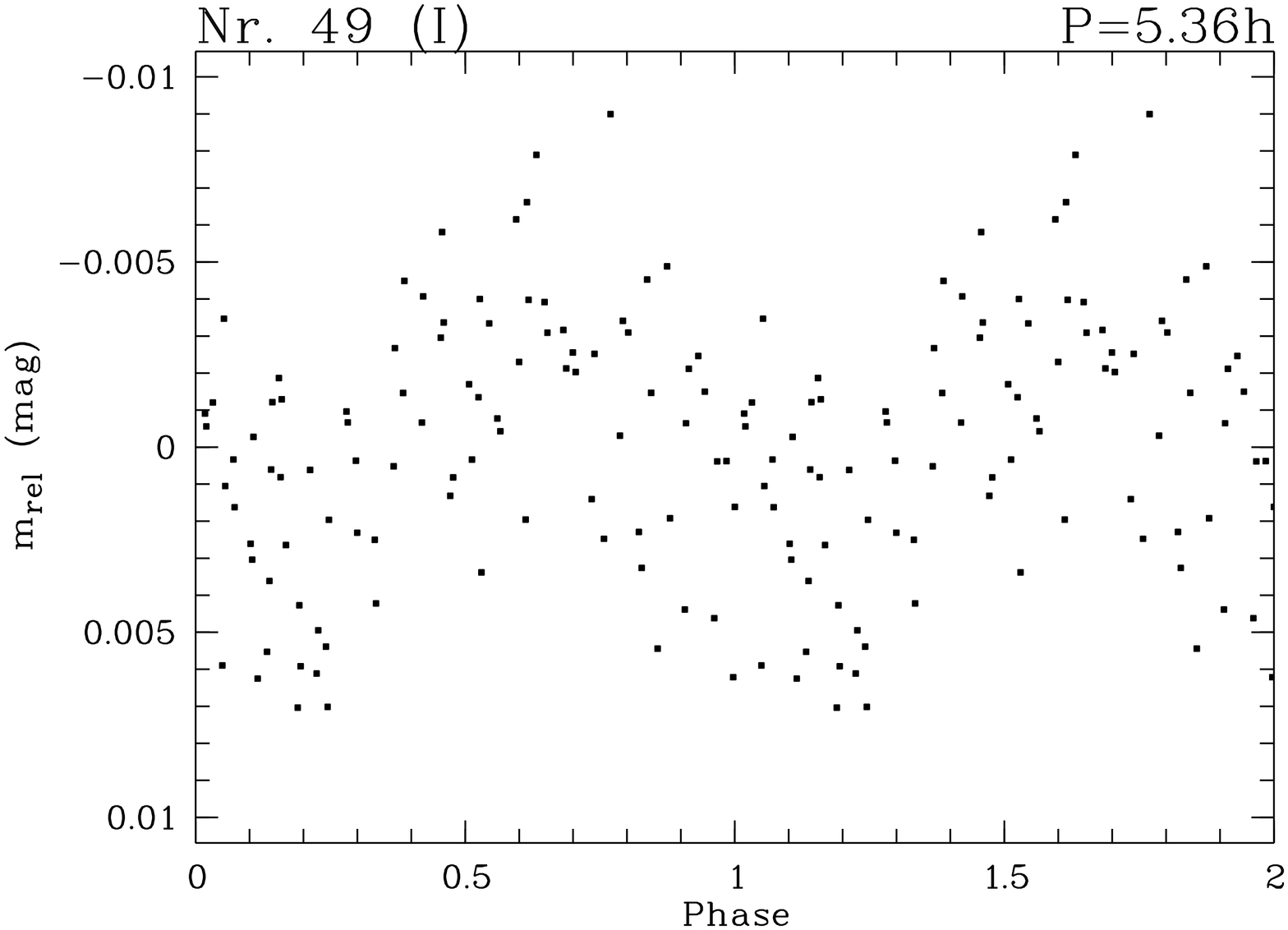}  \hfill
\includegraphics[width=3cm,angle=-90]{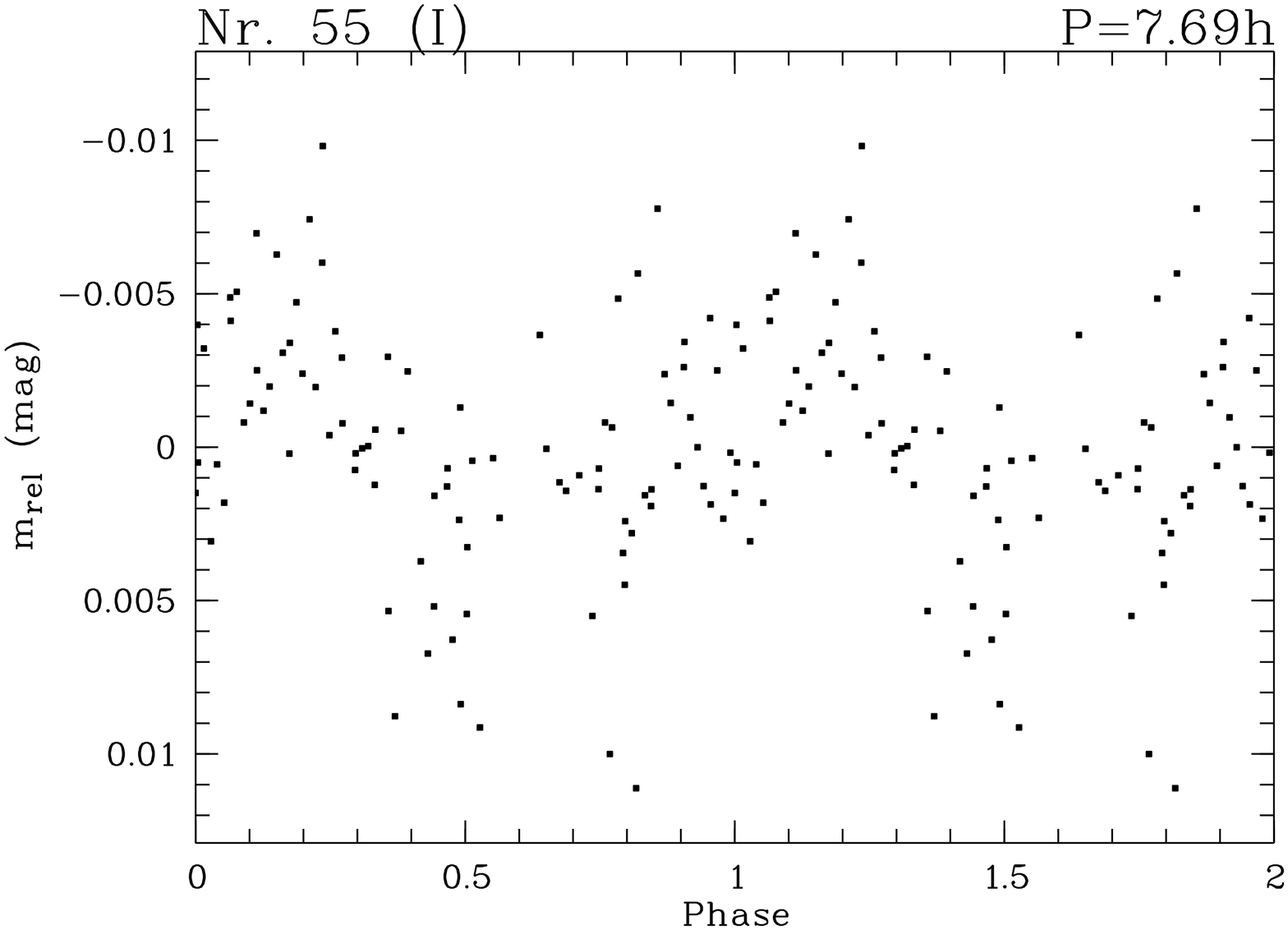} \hfill
\includegraphics[width=3cm,angle=-90]{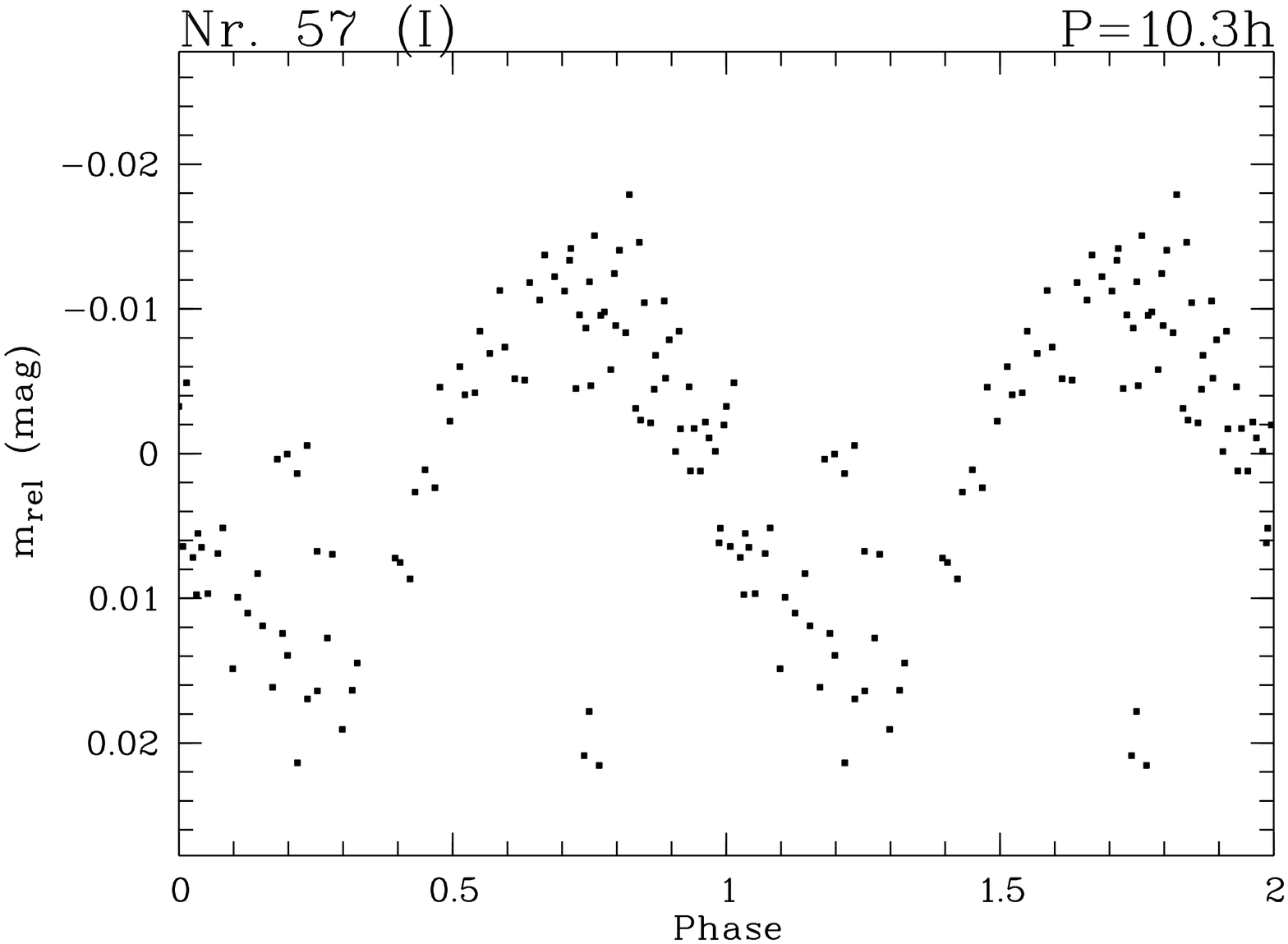} \\
\includegraphics[width=3cm,angle=-90]{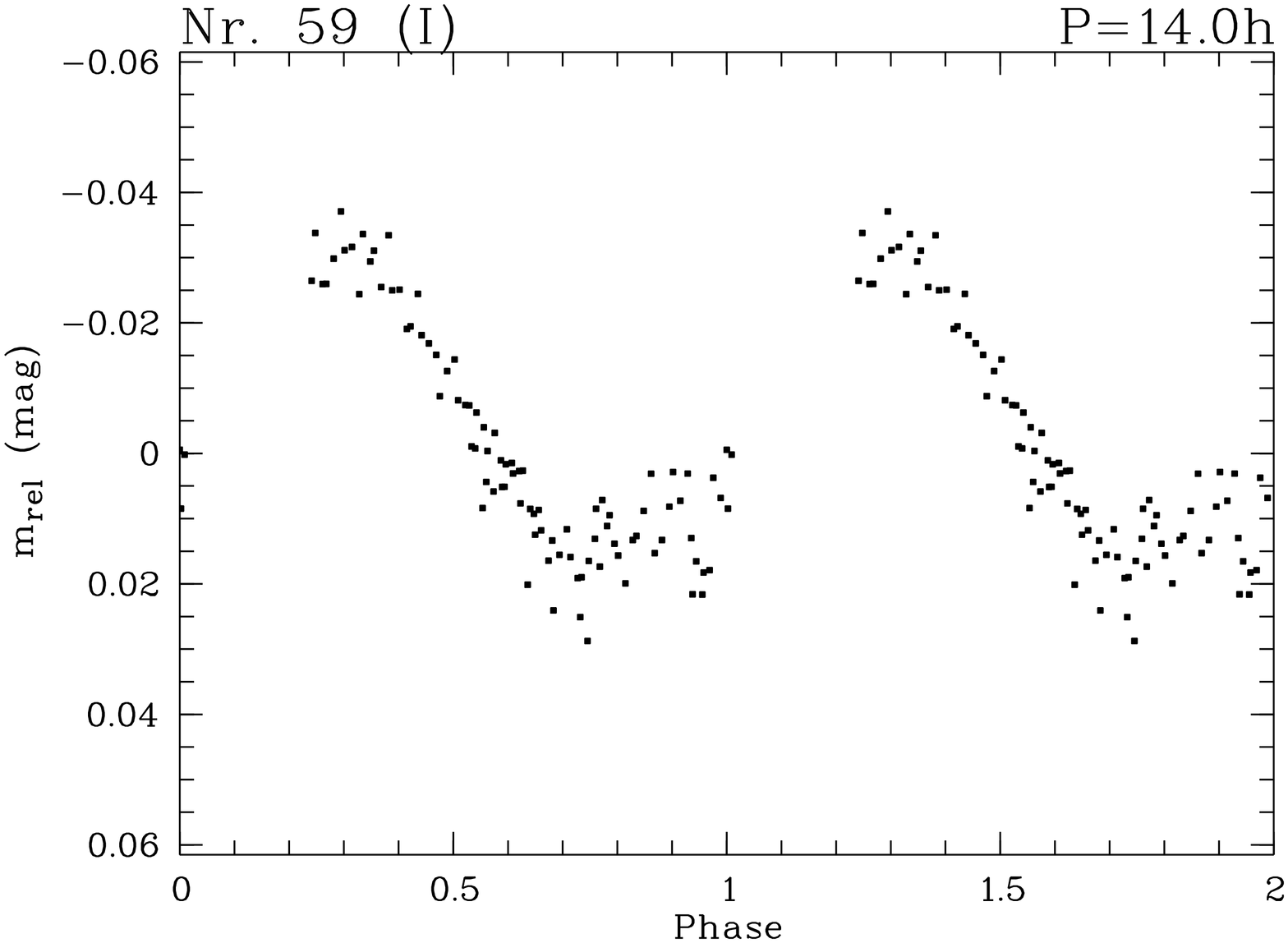} \hfill
\includegraphics[width=3cm,angle=-90]{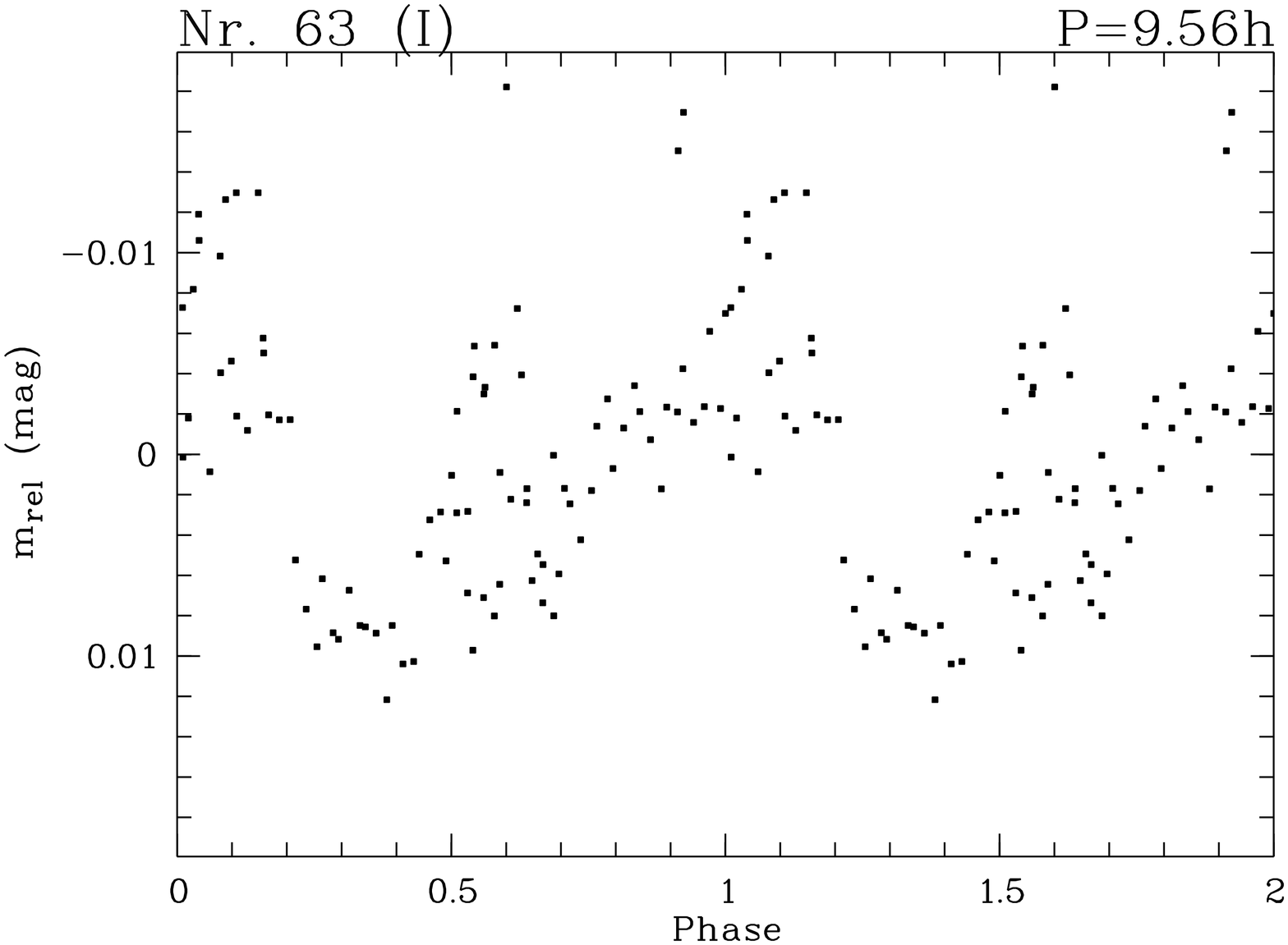} \hfill
\includegraphics[width=3cm,angle=-90]{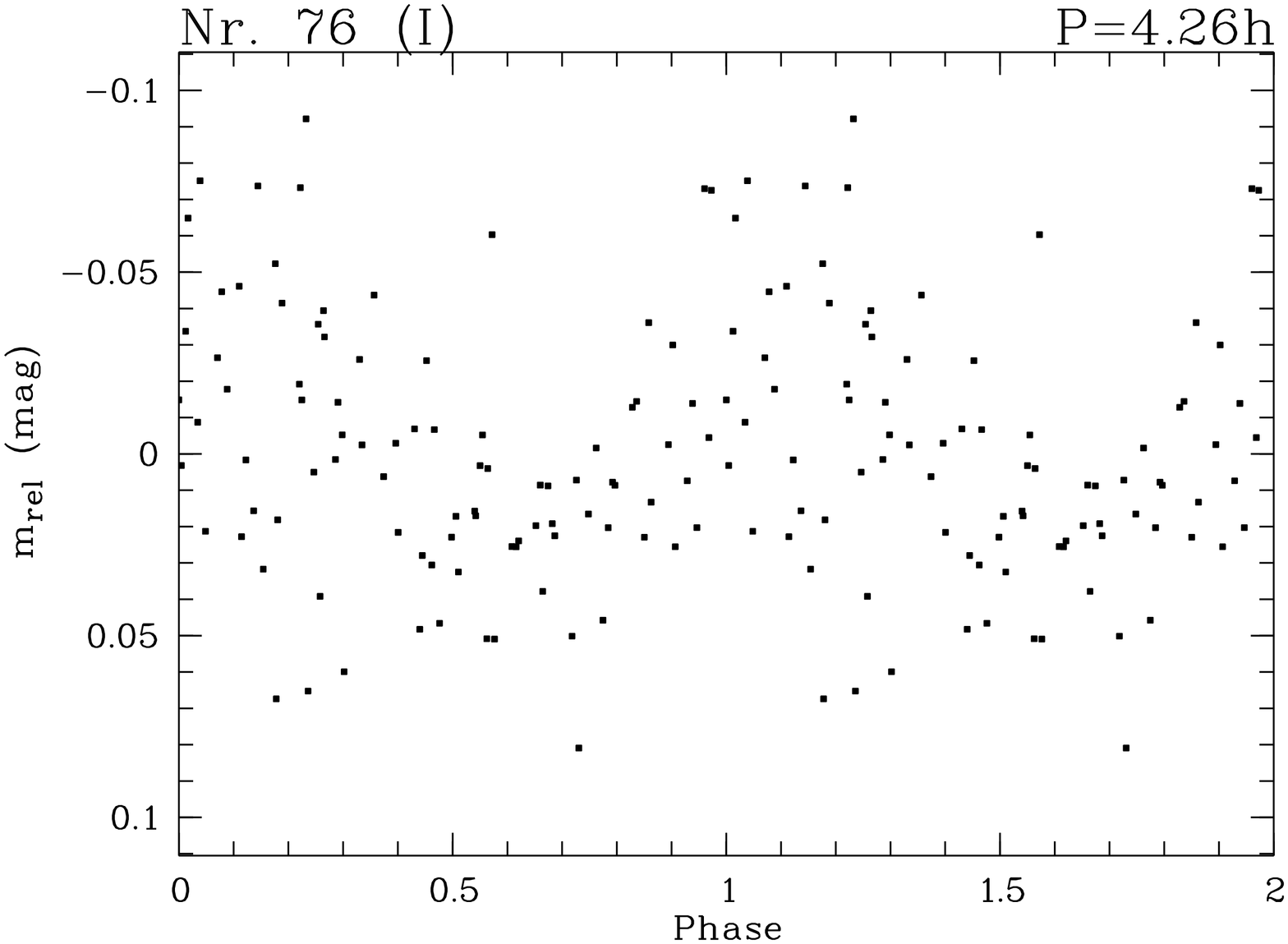} \hfill
\includegraphics[width=3cm,angle=-90]{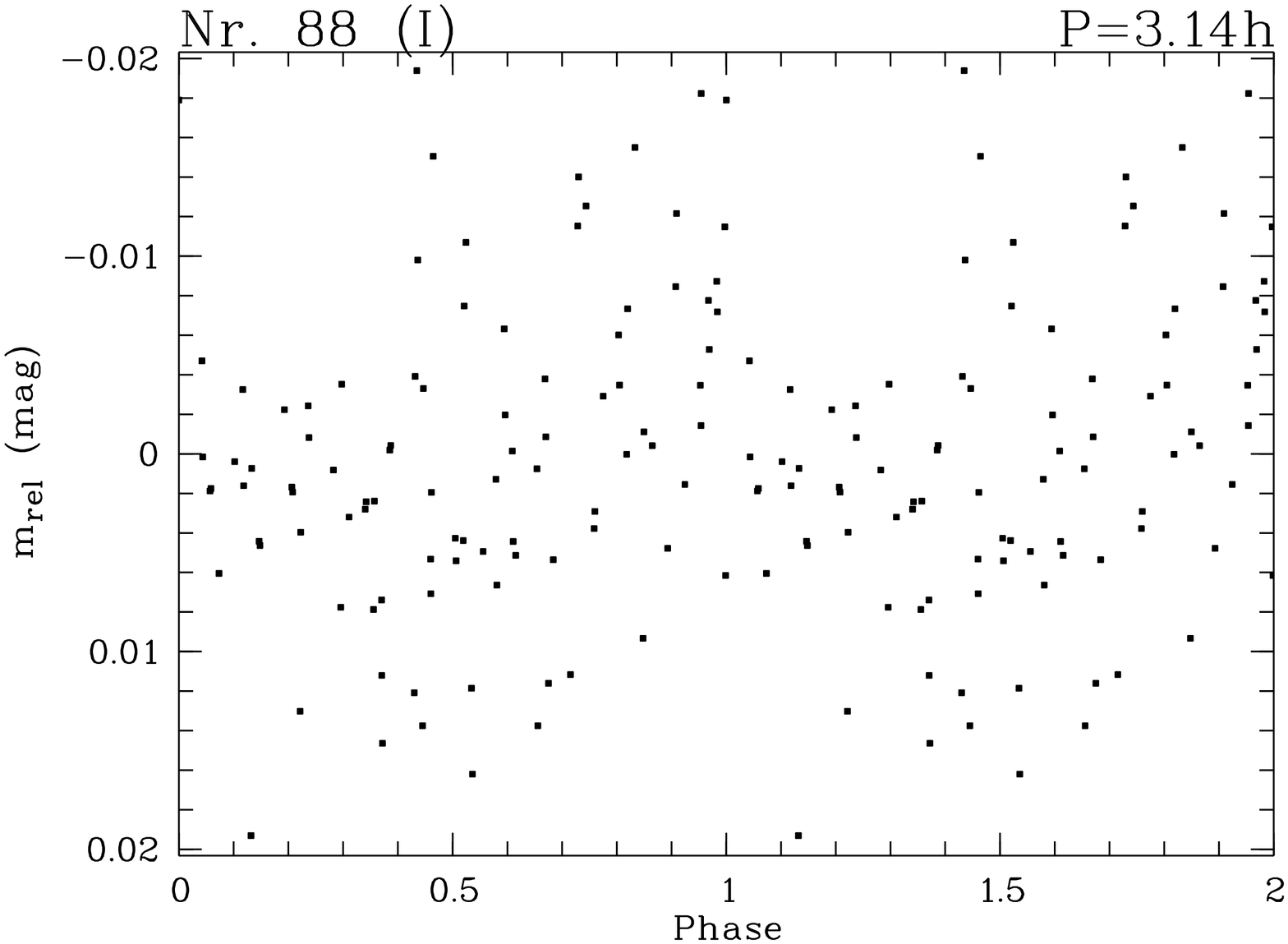}  \\
\includegraphics[width=3cm,angle=-90]{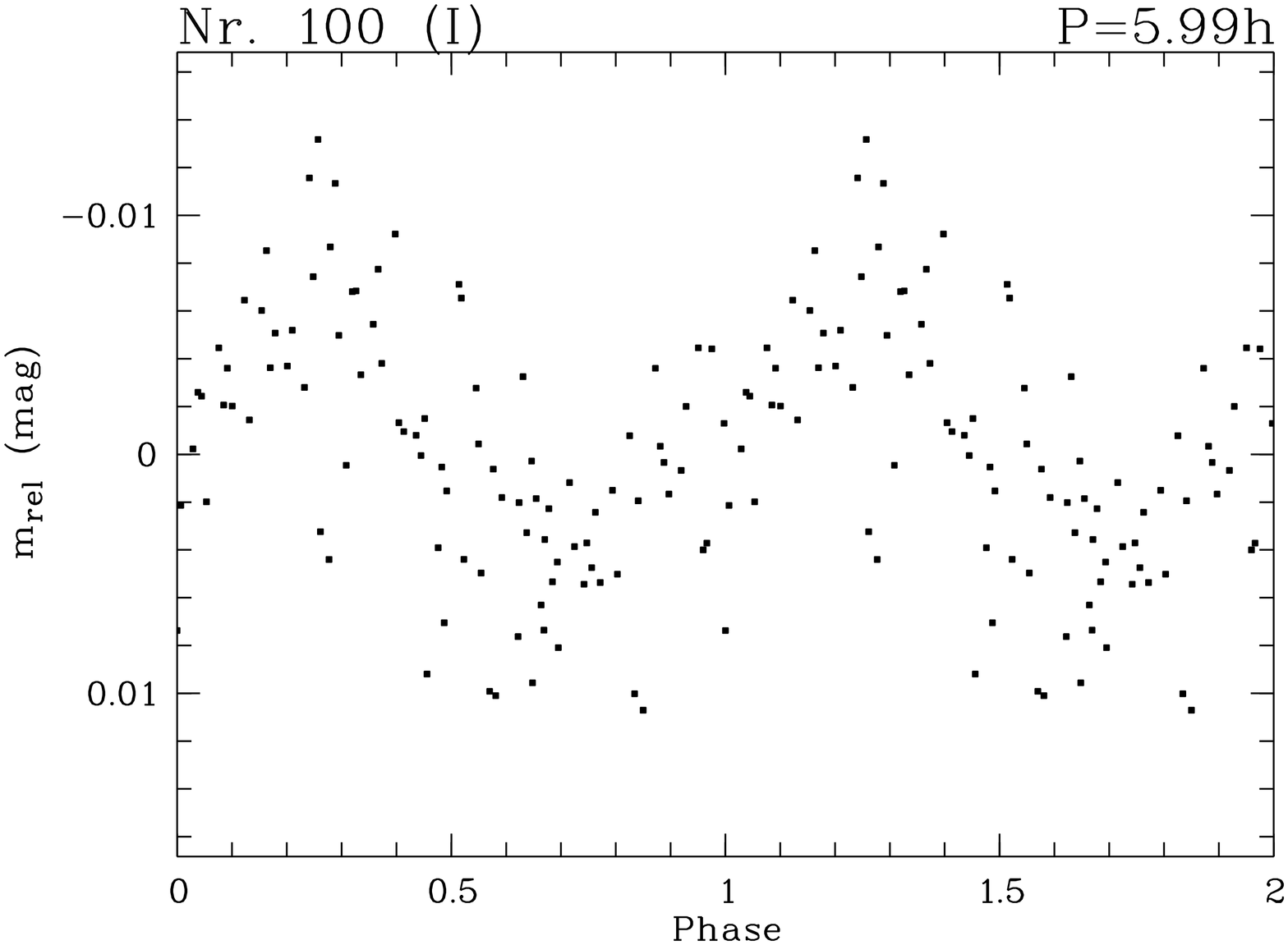} \hfill
\includegraphics[width=3cm,angle=-90]{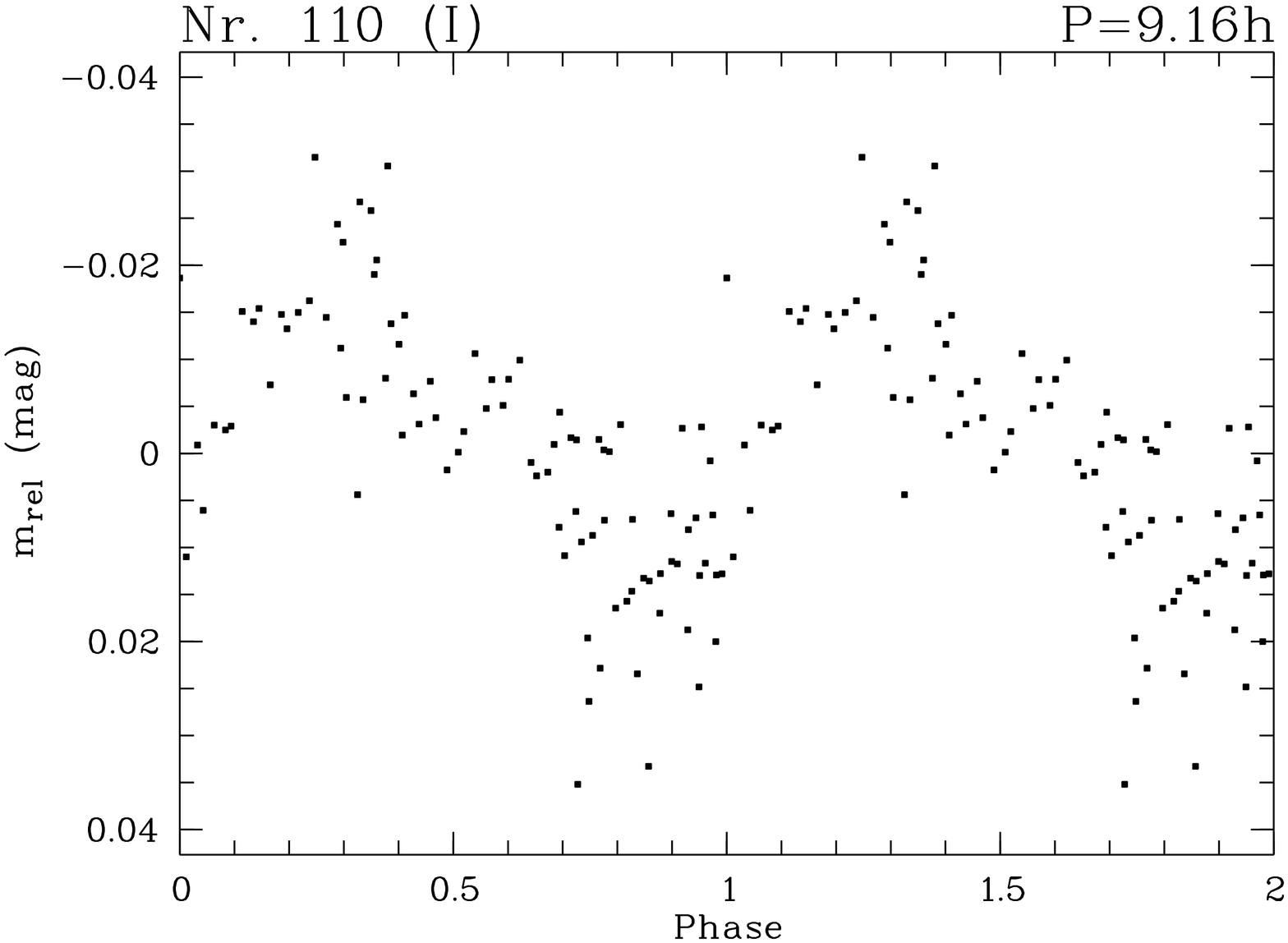} \hfill
\includegraphics[width=3cm,angle=-90]{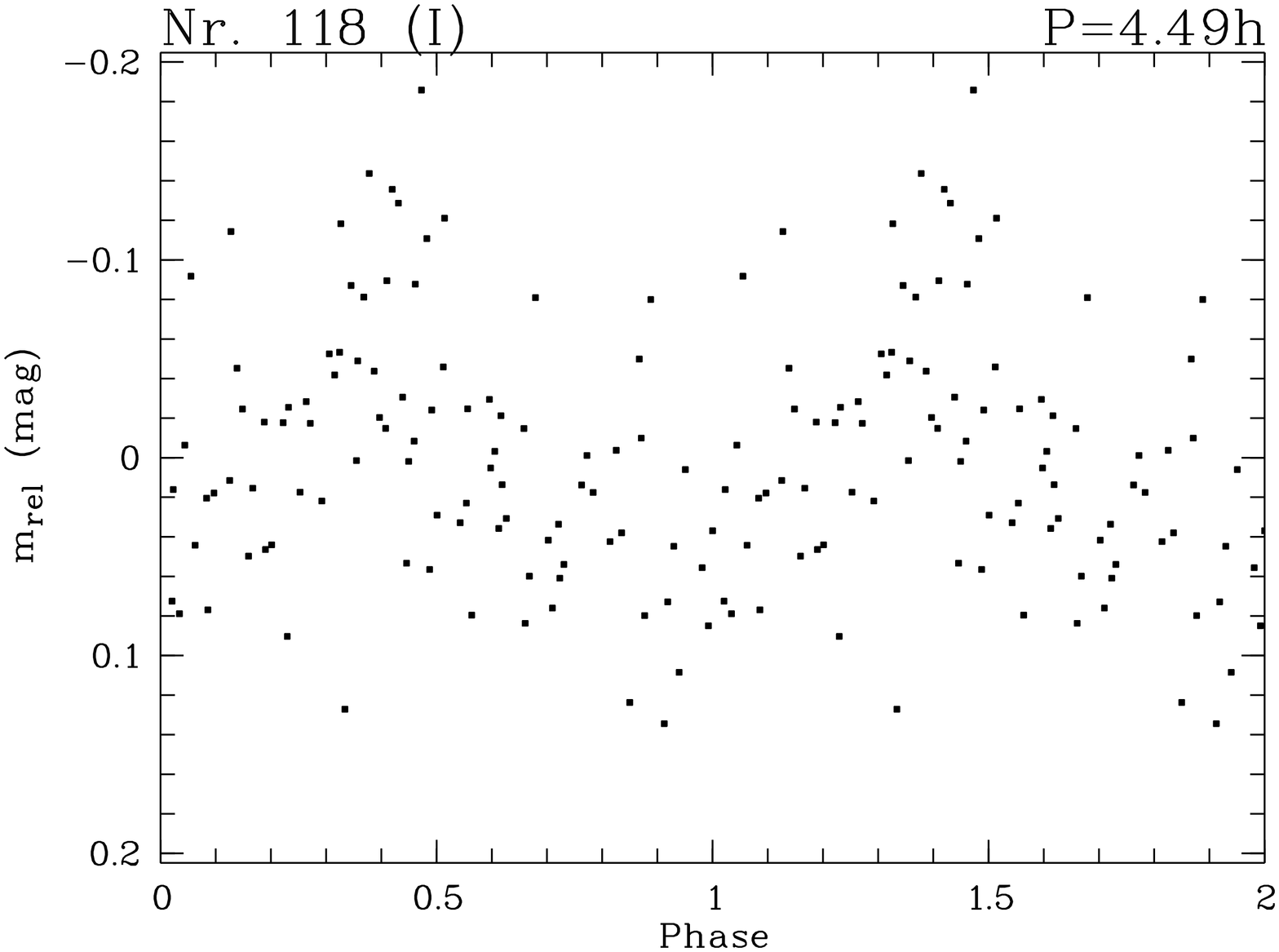} \hfill
\includegraphics[width=3cm,angle=-90]{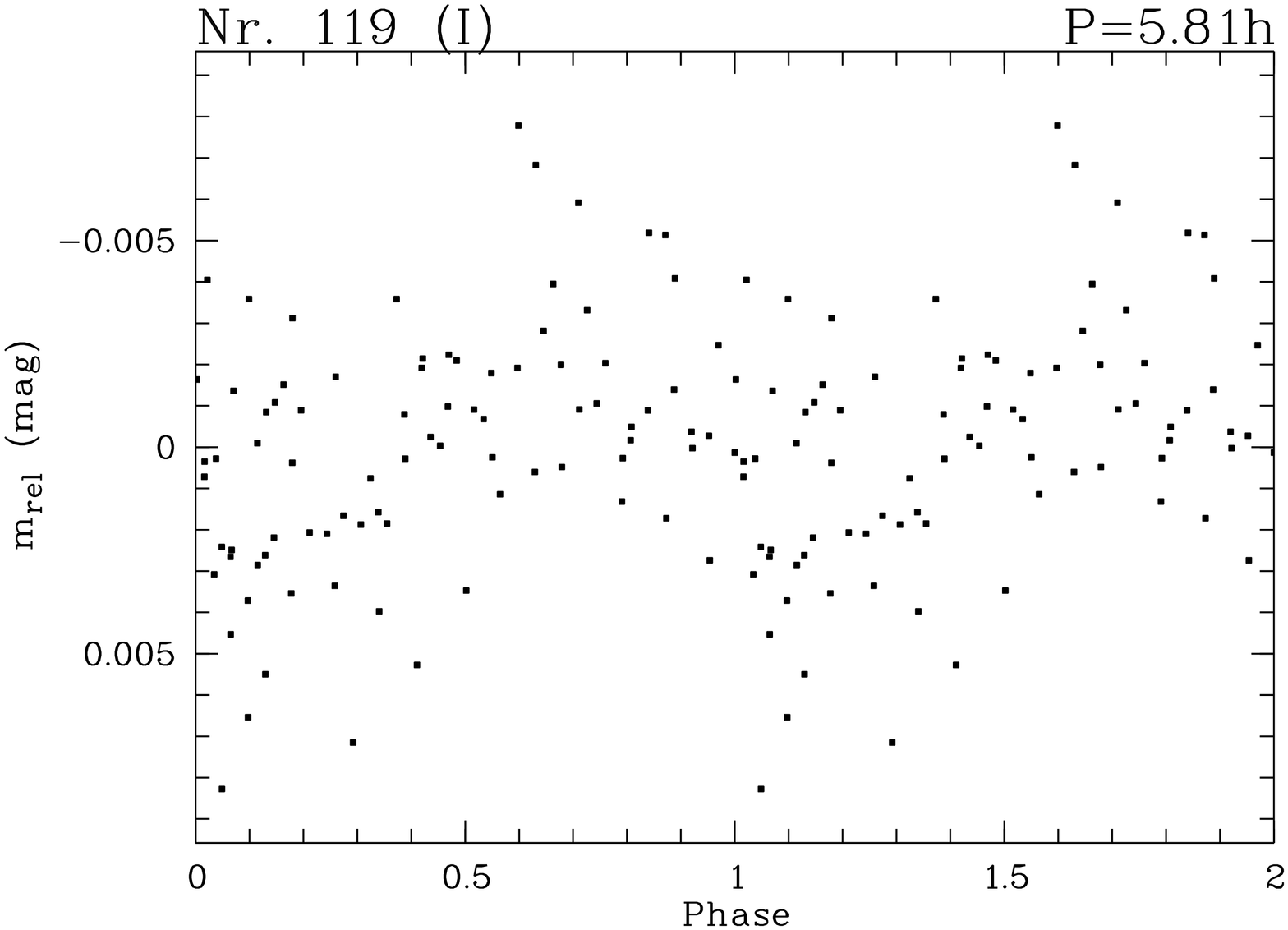} \\
\caption{Phased lightcurves for the periodic objects from run A. Object id and photometric
period are indicated. \label {f4}}
\end{figure*}

From the 113 VLM members in IC4665, 16 show significant periodic variability in run A and 6 in
run C. Only two of these objects exhibit a photometric period in both campaigns. Final false alarm
probabilities for the periodicities are estimated using the 'bootstrap' approach: For each candidate,
a set of 10000 random lightcurves was generated by shuffling the photometry datapoints, while retaining
the time sampling. The highest peak in the Scargle periodogram was determined for all these test
lightcurves, and compared with the peak height found for the periodicity in the actual lightcurve.
A FAP$<0.01$\% indicates that none of the 10000 random time series resulted in a Scargle peak higher than
the one in the observed lightcurves. All our periods turn out to be highly significant according to
this test. Table \ref{periods} lists the derived periods and further results from the time series
analysis. The period errors $\Delta P$ in the table are determined following \citet{1986ApJ...302..757H}; 
for comparison we also derived $\Delta P$ from the FWHM of the peak in the Scargle periodogram
which typically gives 1-3\% uncertainty. The amplitudes in the table correspond to the peak-to-peak 
amplitude of a sinefit to the lightcurve, not the lightcurve itself, to avoid being affected by the 
noise in the photometry.

\begin{table}
\caption{Results of the period search: objects with significant periodic
variability in runs A and B. (id: identification number from Table \ref{cands});
P: period; $\Delta$P: period error; A: amplitude (see text); FAP: false alarm 
probability (see text); N: number of datapoints used in the time series 
analysis}
\label{periods}
\begin{tabular}{lccccc} 
\hline
ID      & P (h) & $\Delta$P (h) &A (mag) & $\mathrm{FAP}$ (\%) & N \\
\hline
{\bf Run A:}\\
%\noalign{\smallskip}
~~~7  &  3.64 & 0.08 & 0.009 & ~~0.03    & 91 \\ %5 4803	 
~~31  &  2.99 & 0.02 & 0.019 & $<0.01$ & 91 \\   %3 4858	
~~36  &  5.45 & 0.01 & 0.045 & $<0.01$ & 93 \\ %  6 4045	 
~~37  &  6.72 & 0.02 & 0.027 & $<0.01$ & 89 \\   %6 1418	
~~40  &  16.5 & 0.08 & 0.047 & $<0.01$ & 94 \\   %3 3263	 
~~49  &  5.36 & 0.08 & 0.006 & $<0.01$ & 90 \\   %3 708	 
~~55  &  7.69 & 0.06 & 0.007 & $<0.01$ & 92 \\   %7 1946	
~~57  &  10.3 & 0.06 & 0.022 & $<0.01$ & 95 \\   %2 2943	
~~59  &  14.0 & 0.04 & 0.045 & $<0.01$ & 93 \\   %2 3867	
~~63  &  9.56 & 0.23 & 0.015 & $<0.01$ & 88 \\   %7 4352	
~~76  &  4.26 & 0.02 & 0.052 & ~~0.05    & 92 \\ %2 4919	
~~88  &  3.14 & 0.02 & 0.012 & ~~0.07    & 93 \\ %1 634	
~100  &  5.99 & 0.02 & 0.010 & $<0.01$ & 93 \\   %8 1409	
~110  &  9.16 & 0.04 & 0.033 & $<0.01$ & 93 \\   %1 2507	
~118  &  4.49 & 0.06 & 0.122 & $<0.01$ & 92 \\   %1 4112	
~119  &  5.81 & 0.11 & 0.005 & $<0.01$ & 84 \\   %1 446	
{\bf Run C:}\\
~~28 & 4.56& 0.02 &  0.011& ~~0.03  & 86  \\  %4470 6	
~~39 & 9.20& 0.03 &  0.020& $<0.01$  & 73  \\ %4990 6   
~~57 & 16.3& 0.12 &  0.022& $<0.01$  & 87  \\ %2943 2	 
~~59 & 14.4& 0.04 &  0.072& $<0.01$  & 86  \\ %3867 2	 
~~73 & 27.0& 0.35 &  0.028& $<0.01$  & 87  \\ %5109 2 
~~75 & 4.84& 0.02 &  0.033& $<0.01$  & 87  \\ %2668 2	   
\hline			     
\end{tabular}			     
\end{table}	

An assessment of the completeness of the period search can be made based on the time sampling
in our observing campaigns. The upper frequency limit for the period search, thus the lower period
limit, is frequently given as $\nu_{\mathrm{max}}=1/2\delta$, where $\delta$ is the minimum separation
between two datapoints \citep[e.g.][]{1987AJ.....93..968R}. For our WFI time campaigns, this gives
a minimum period of $P_{\mathrm{min}}=0.4\,h$. Since we did not find any period with $P<3$\,h, we are
complete at short periods. With regular sampling, the longest period than can reliably be detected is 
roughly given by the total length of the time series (from first to last datapoint), which is 12-13 days 
in the WFI runs. Nights with very few datapoints ($\le 5$, see Table \ref{ts}) are unlikely to extend 
the baseline; excluding those nights gives $P_{\mathrm{max}}=1.3$\,d for run A and 6\,d for run C. 

\begin{figure}
\includegraphics[width=2.9cm,angle=-90]{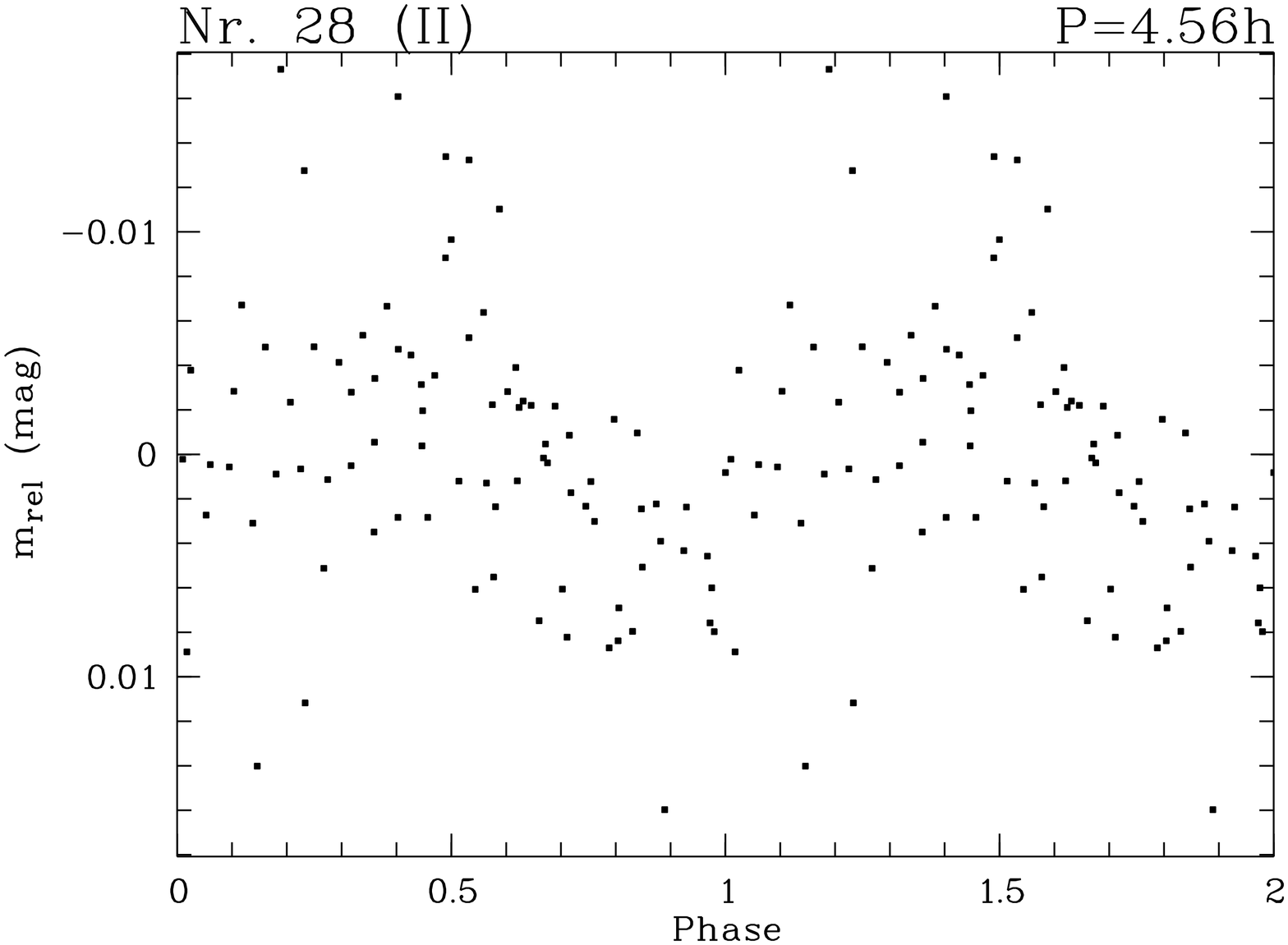} \hfill
\includegraphics[width=2.9cm,angle=-90]{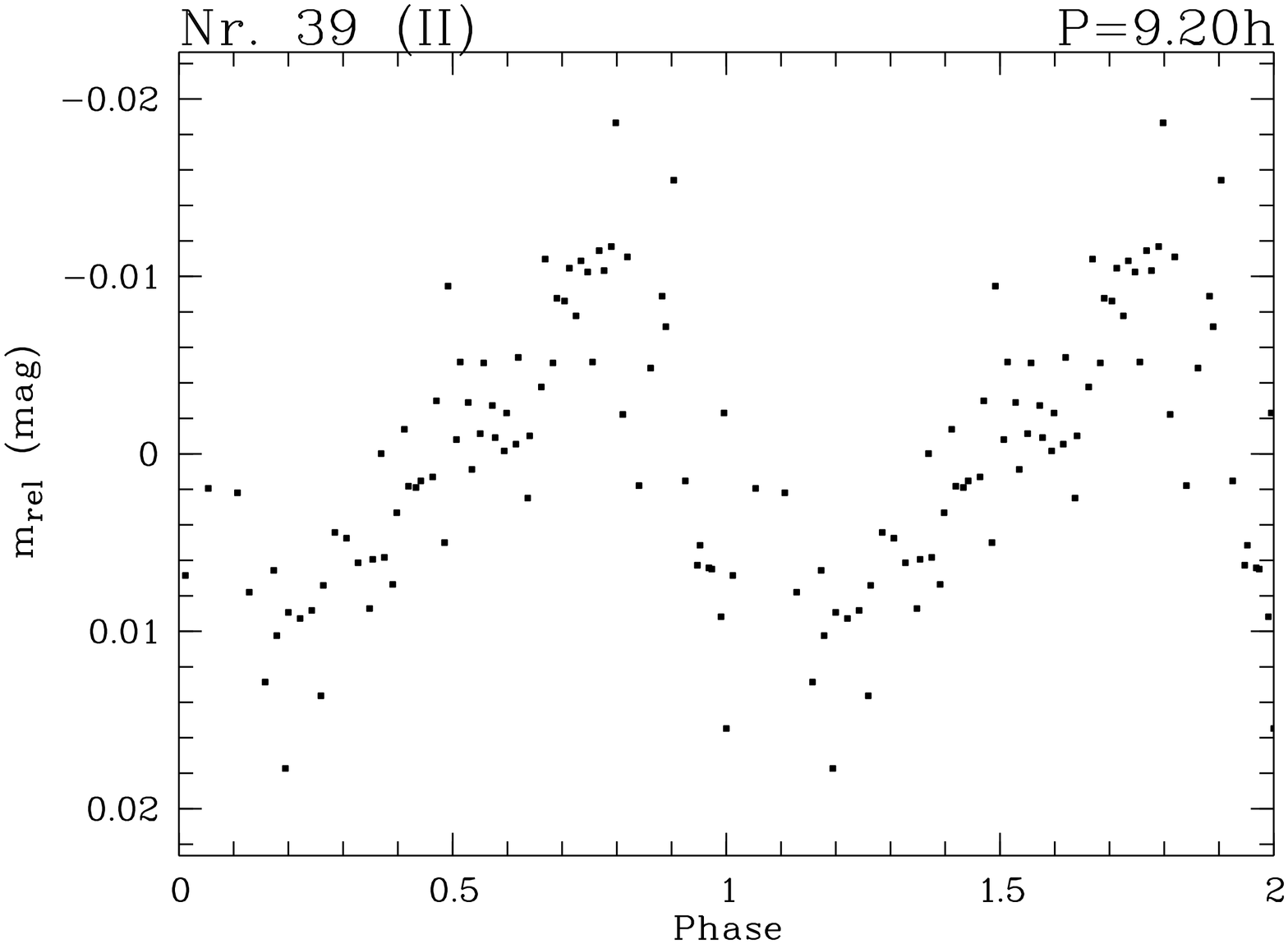} \\
\includegraphics[width=2.9cm,angle=-90]{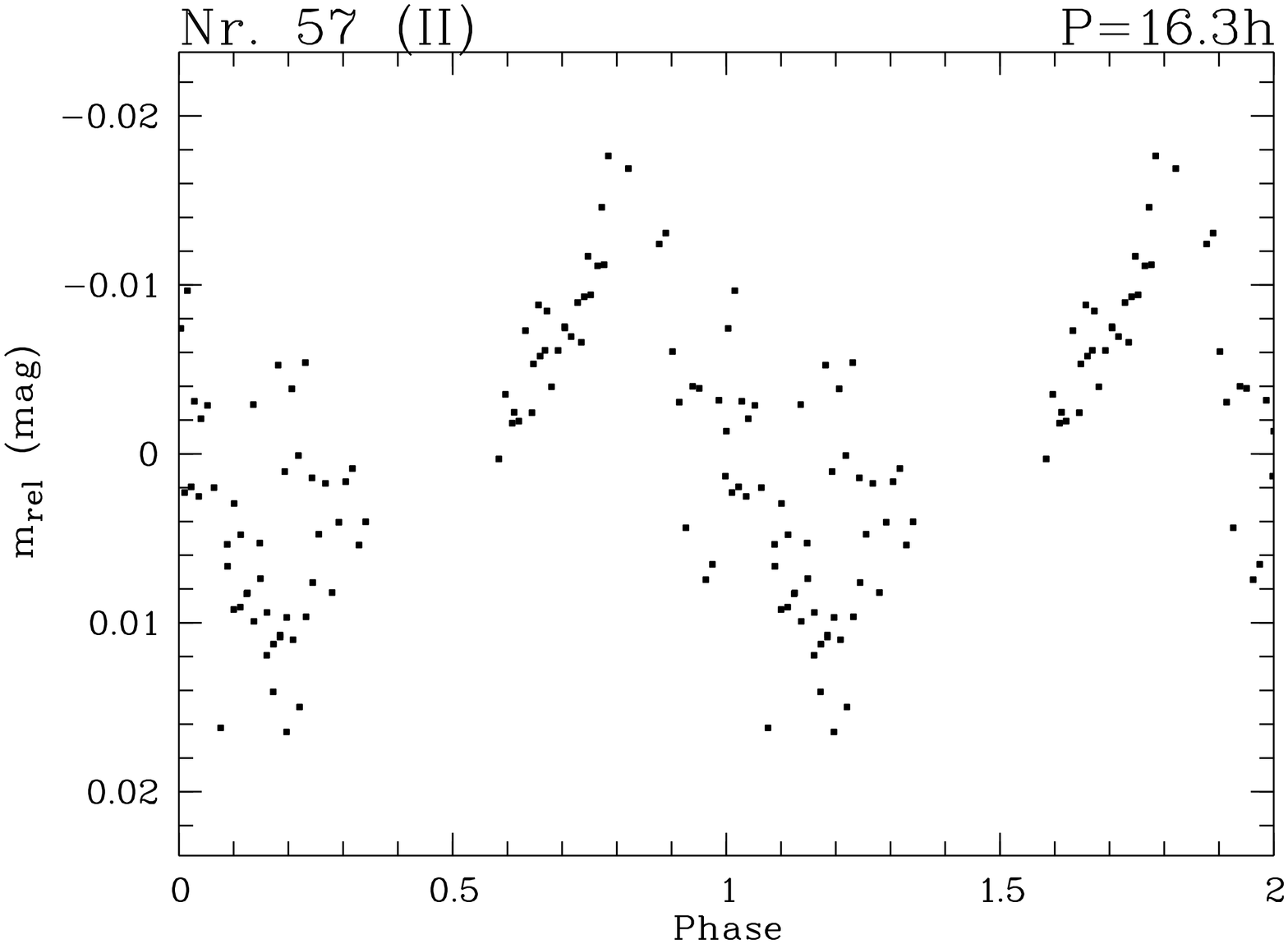} \hfill
\includegraphics[width=2.9cm,angle=-90]{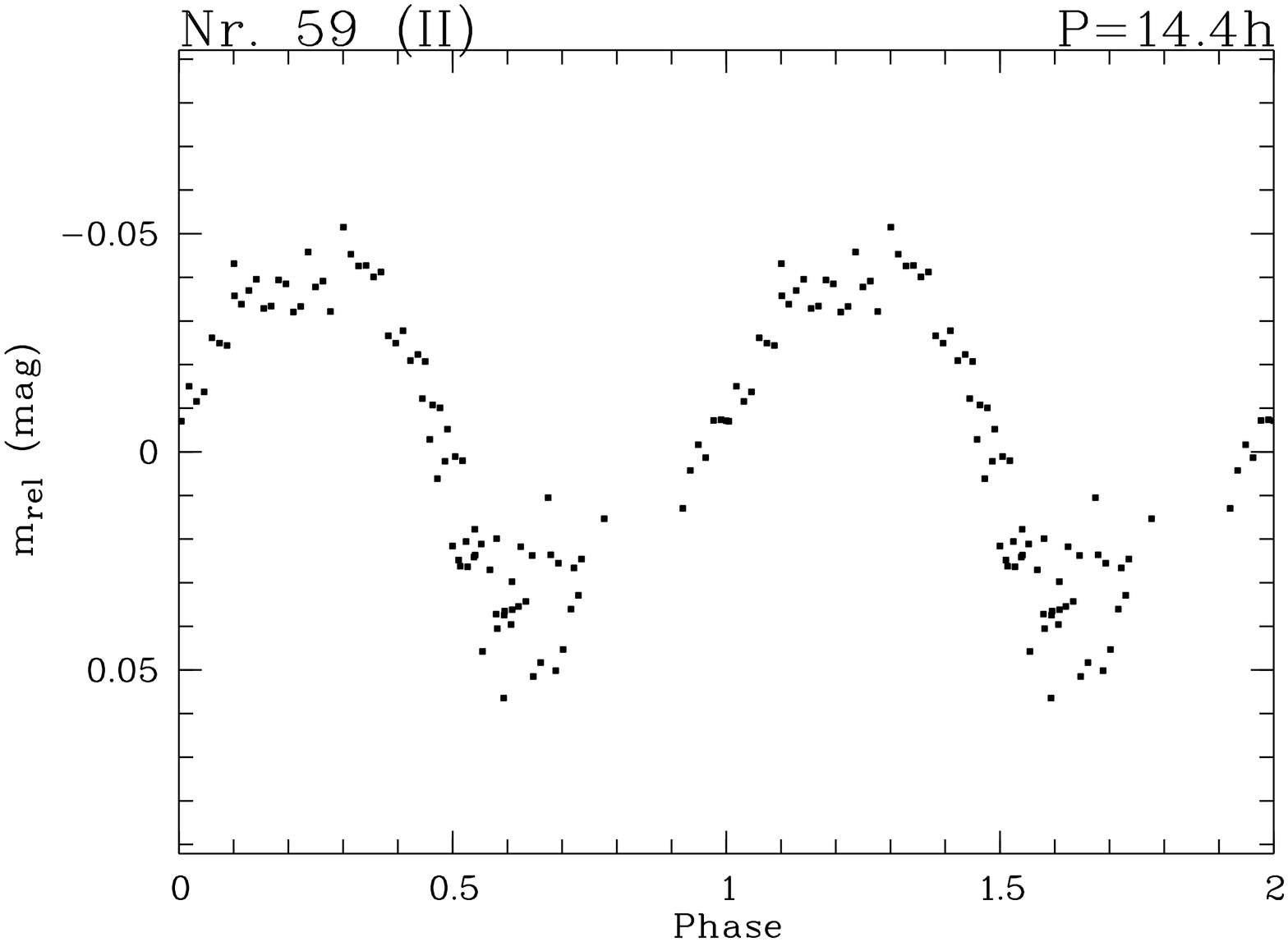} \\
\includegraphics[width=2.9cm,angle=-90]{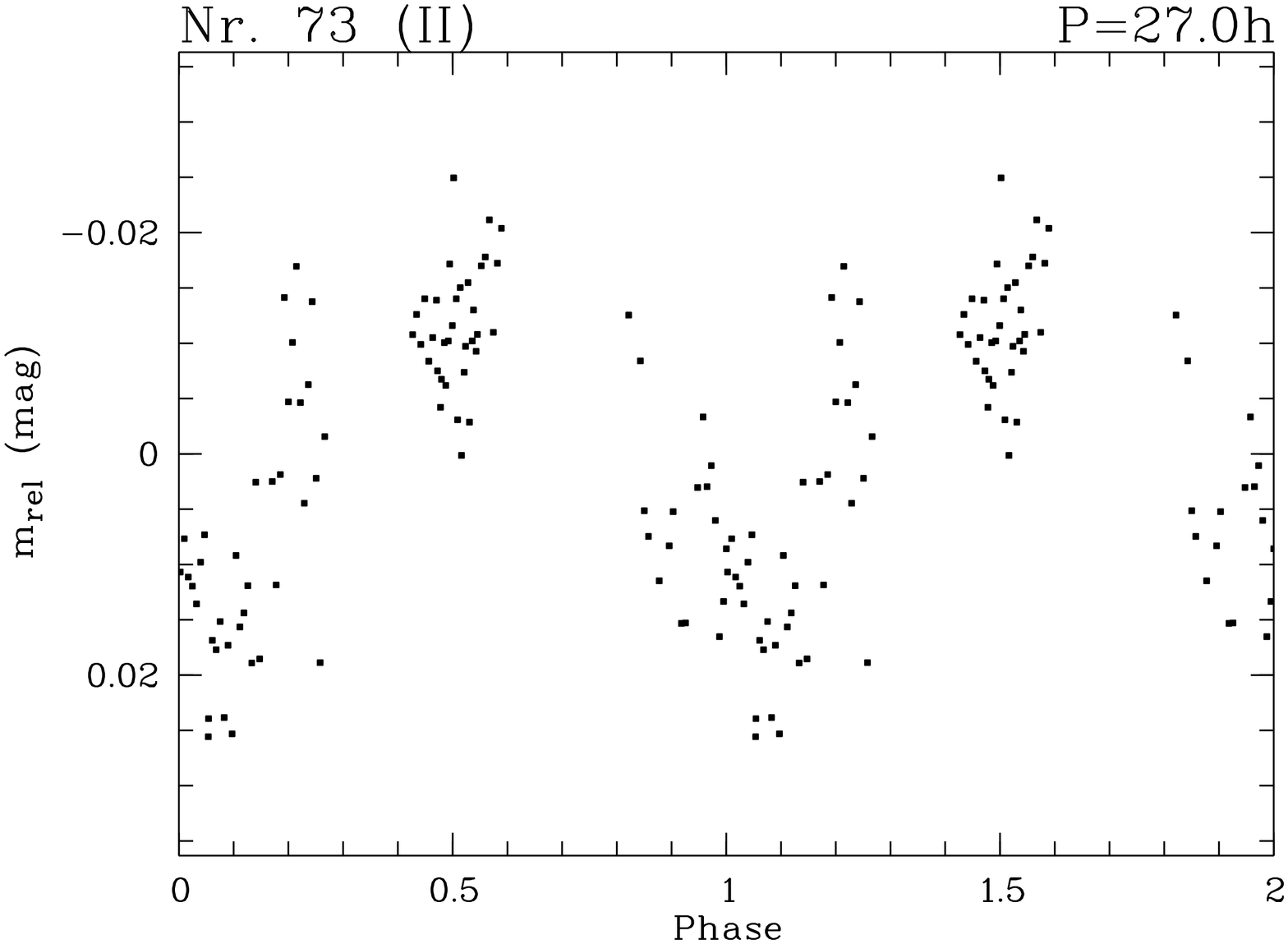} \hfill
\includegraphics[width=2.9cm,angle=-90]{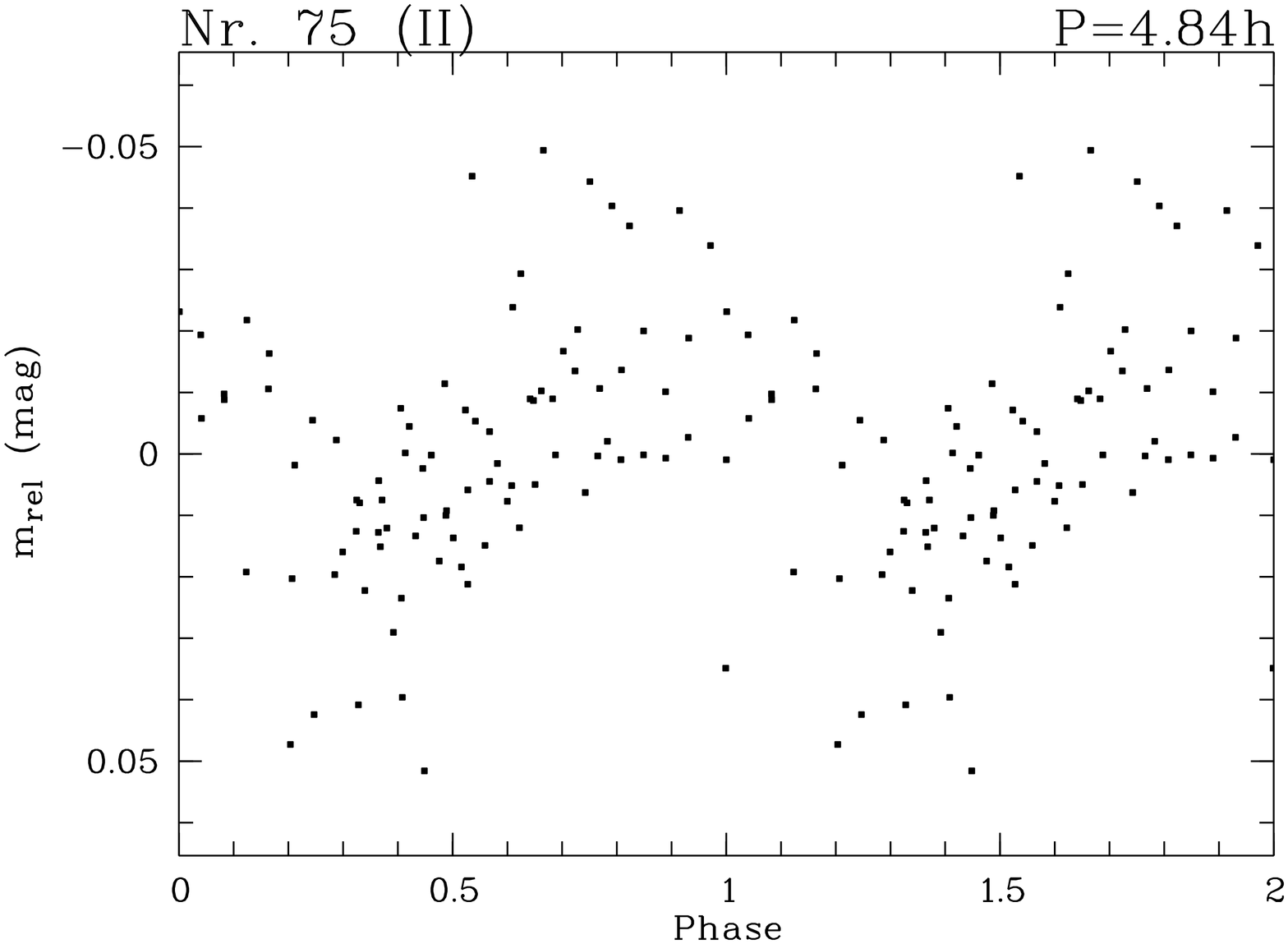} \\
\caption{Phased lightcurves for the periodic objects from run C. Object id and photometric
period are indicated. \label{f5}}
\end{figure}

For a more realistic estimate of the sensitivity, we carried out simulations following the recipe
given in \citet{2004A&A...421..259S}. In short: Non-variable objects were selected from our database.
Sine-shaped periodicities with periods ranging from 1 to 200\,h were added to their lightcurves. 
The noise level of the test objects and the amplitudes of the artificial periods were chosen to 
approximately reproduce the typical properties of the periods listed in Table \ref{periods}. We then
applied the algorithm described in Sect. \ref{perser} to these lightcurves and aimed to recover the 
imposed periods.

\begin{figure}
\includegraphics[width=6cm,angle=-90]{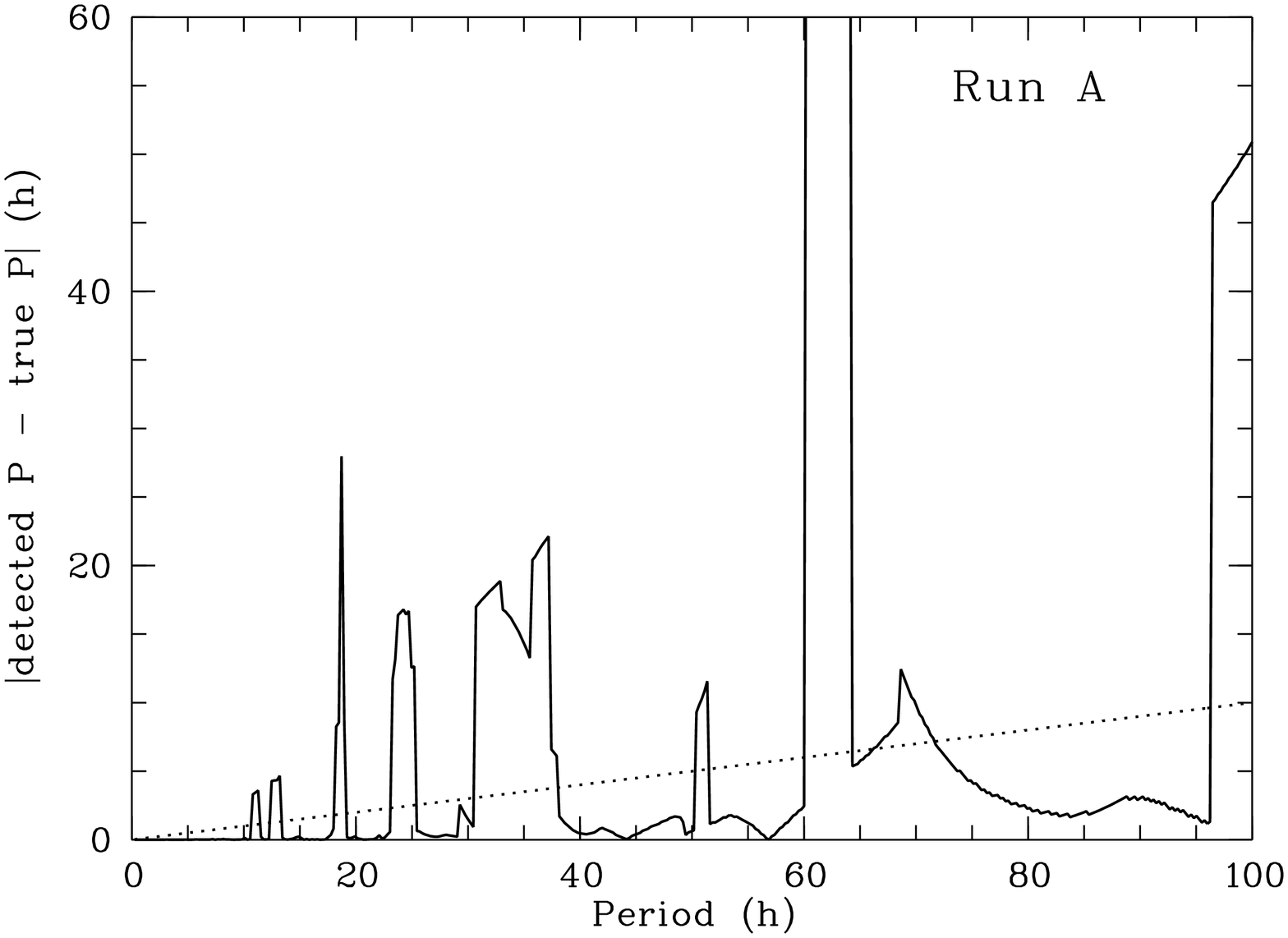}
\includegraphics[width=6cm,angle=-90]{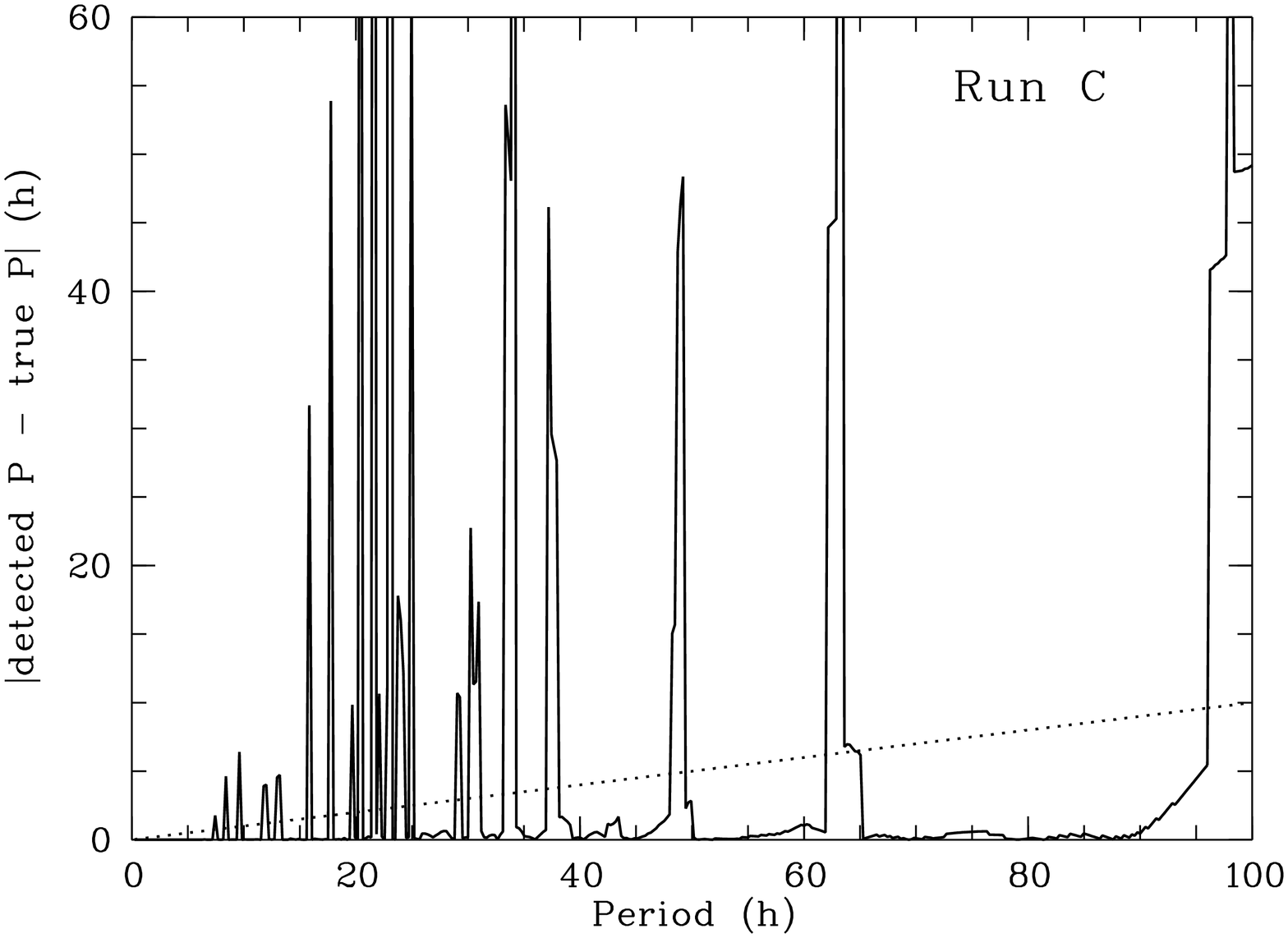}
\caption{Results of the simulations to test the sensitivity in the period search. Plotted is the
difference between imposed and detected period vs. the period for runs A (upper panel) and C
(lower panel). The dotted line
corresponds to a deviation of 10\%.\label{f12}}
\end{figure}

Fig. \ref{f12} shows the results fo these simulations for runs A and C. The deviation between the imposed 
period and the detected period is negligible for $P<10$\,h, confirming that our period search is optimal 
for this range of periods. For $P=10 \ldots 90$\,h there are certain period ranges (e.g., around 24\,h) 
for which we are unable to obtain reliable detections. In many cases, these ranges are narrow 
(typically $\delta P<0.5$\,h), particularly for run C. The plots clearly demonstrate that there are 
wide ranges of periods $>10$\,h in which our time series analysis is sensitive. For example, 
in run A we recover most periods with $P=40\ldots 60$\,h, and in run C we reliably recover 
$P=50\ldots 62$\,h and $P=65\ldots 90$\,h. There are only negligible windows of sensitivity for $P>95$\,h, 
which we identify as our upper period detection limit. The non-existence of any significant periods 
with $P>30$\,h in our candidate lightcurves thus points to a genuine lack of such objects in IC4665 
(see Sect. \ref{rotation}).

\subsection{The TLS lightcurves (run B)}
\label{tls}

While the photometry from runs A and C comes with similar noise characteristics, run B suffers from 
a comparatively small dynamic range and increased noise level, see Fig. \ref{f1}. This hampers a direct 
comparison between the three runs. Instead of carrying out a time series analysis for all 
candidate lightcurves from run B, we use these time series to verify the periods reported in Table
\ref{periods}. 

Out of 20 objects which are periodic in at least one of the WFI runs, 15 are bright enough to be
detected in the TLS images. Nine out of these 15 do not show a significant period according to our
tests. For the two objects \#55 and \#39 we confirm the periods listed in Table \ref{periods} within
$\pm 0.1$\,h. For two objects we find periods at odds with the ones detected in the WFI runs:
Object \#63 has a 7.7\,h period (run A: 9.6\,h), object \#100 a $\sim 110$\,h period (run A: 6.0\,h).
The latter case could indicate a long-term trend in the lightcurve on timescales of $>4$\,d and thus
not detectable in the WFI runs. For both ambiguous cases, we re-checked the WFI periods and found them 
to be more convincing. 

Of particular interest are the two objects for which both WFI runs yielded a period detection. For 
candidate \#57 the best periods from the WFI campaigns are 10.3\,h in run A and 16.3\,h in run C. In
both cases, however, the periodograms (Scargle and CLEAN) do exhibit a weaker, but still significant 
peak at the period found in the other run, respectively. The two possible periods are neither 
harmonics ($P_1 = n P_2$) nor beat periods ($1/P_1 = 1/P_2 + n$). In run B, this object is again variable 
with a period of 18.4\,h, which is a clear beat period of 10.3\,h, the period from run A (with $n = -1$). 
Thus, $P=10.3$\,h is the only period consistent with all three time series and is adopted as the most 
likely rotation period for this object in the following sections. The fact that the most significant
period in dataset C differs from our consensus period, illustrates the difficulties of assigning periods
to individual objects based on a single monitoring run. Due to the incomplete sampling in ground-based 
data, the lightcurve for a given object can sometimes be fit by more than one convincing period. This
can be overcome with multiple observing runs, as demonstrated here.

Object \#59 shows a period of 14.0\,h in run A and 14.4\,h in run C. The candidate is clearly variable 
in run B: Its lightcurve shows a gradual brightening of $\sim 0.04$\,mag over the five nights. In 
addition, in night 5 the lightcurve exhibits several outlying datapoints which are 0.05-0.2\,mag 
brighter than the average trend, but do not show the typical characteristics of a flare (see Sect. 
\ref{tsa}). For comparison, the average noise level for objects with similar brightness
is 0.015\,mag. However, we do not find any significant period in the lightcurve. In particular, the 
'consensus' period of $\sim 14$\,h obtained from the WFI lightcurves is not detected. Thus, while 
the variability of this object is confirmed, the period is not.

\section{Magnetic spots on very low mass objects}
\label{spots}

We found periodic variability for 20 likely VLM members of IC4665. The best explanation for
this type of variability is the presence of surface features, co-rotating with the objects and thus 
modulating the flux, as it is usually assumed in monitoring studies for this type of targets. Alternative
interpretations include pulsations and eclipsing binaries, but both are unlikely for our sample: a) Eclipsing
binaries are simply too rare to be responsible for a significant fraction of our periodic objects. b)
There is no evidence which points to pulsation in young M dwarfs with the periods observed in this study.
All further discussions are based on the assumption that with the perodicities we indirectly observe surface 
features. 

In our observations, the surface features causing the periodicities are almost certainly magnetically 
induced cool spots, simply for the lack of an alternative plausible scenario. It has been suggested that 
the condensation of dust molecules will form 'clouds' in substellar objects, which could in principle
account for periodic variability as well. Our young targets, however, are not yet cool enough for this 
to happen. Models predict that dust condensation becomes relevant at $T_{\mathrm{eff}}< 2500$\,K 
\citep{2008A&A...485..547H}, while the coolest objects in the sample given in Table \ref{cands} are 
likely to have $T_{\mathrm{eff}}>2700$\,K. Moreover, our targets are expected to be mid to 
late M dwarfs, a spectral range known for strong magnetic activity. Thus, by analysing the lightcurves
from our monitoring campaigns, we can put limits on properties of magnetic spots on VLM objects.

It is useful to recapitulate the conditions required to detect a photometric rotation period. Amplitudes 
and lightcurve shape will be influenced by the distribution of the spots on the surface. A degree of 
asymmetry in the distribution is required to cause a photometric period, but a slight deviation from 
symmetry would be sufficient to explain the small amplitudes seen in this study. Moreover, low-latitude 
spots are more relevant than polar spots for the lightcurve appearance.

The spot coverage (usually measured as filling factor) and temperature contrast (difference between photospheric 
and spot temperature) affect the amplitude of the photometric variability and thus the 'success rate'
in monitoring campaigns. As shown in \citet{2005A&A...438..675S}, both parameters are degenerate and cannot
be determined separately based on single-filter lightcurves. Under the limiting assumption that the spot 
temperature is $T=0$\,K, we can define a lower limit for the spot coverage: In that case, amplitudes of 
1-10\%, as seen in our period sample, would require a spot filling factor of at least 1-10\%. In reality, the 
temperature contrast between photosphere and spots is more likely to be in the range of 100-1000\,K 
\citep{2005A&A...438..675S}, implying more spot coverage than in the limiting $T=0$\,K case.

Finally, the inclination of the rotation axis against the line of sight will affect the lightcurve. For objects 
seen face-on, we expect flat lightcurves, independent of the spot properties. Assuming that the orientation of 
rotation axes in open clusters are randomly distributed, this will lower the fraction of objects with observed 
periodicity by a constant factor.

\subsection{Spot activity vs. mass}
\label{spotmass}

The periods measured in Sect. \ref{perser} provide us with two indicators to assess the level of spot
activity in our sample, the amplitudes and the fraction of objects with periods. The photometric amplitudes 
found in our period sample are $\le 0.05$\,mag in the clear majority of the cases (20 out of 22), 
indicating that VLM objects in IC4665 show only low-level photometric variations. This is also reflected 
in the rms of their lightcurves, as discussed in Sect. \ref{tsa} (see Fig. \ref{f1}). Assuming that the
contamination rate is similar in the total candidate sample and in the periodic sample, the fraction of 
objects with periods is 18\% (20 out of 113). This number increases to 25\% (20 out of 80), if we assume
that the periodic objects have zero contamination (see Sect. \ref{targets}).

These two results will now be compared with the outcomes of variability studies for more massive stars 
to test if spot activity changes with object mass. As comparison samples we mostly focus on four clusters with
similar age, IC4665 itself, IC2602, $\alpha$\,Per, and the Pleiades. Monitoring campaigns in these clusters 
were mostly carried out in the V-band, instead of the I-filter we have used. Assuming typical properties of 
cool spots, the photometric amplitudes are likely to be a factor of 1.5-2 larger in the V-band 
\citep{2005A&A...438..675S}. This has to be taken into account when comparing amplitudes from different 
authors.

In IC4665 \citet{1996A&A...305..498A} monitored 15 F-K dwarfs and find 8 with period with amplitudes 
0.01-0.1\,mag, a 'detection rate' of 53\%, significantly higher than in our campaigns.  
\citet{1999ApJ...516..263B} observed all spectroscopic members of IC2602 and measured periods for 29 out
of 33 F-K dwarfs, a fraction of 88\%. With one exception, all amplitudes are $\le 0.08$\,mag. In 
the $\alpha$\,Per cluster \citet{1993PASP..105.1407P} and \citet{1993MNRAS.262..521O} find photometric 
periods for $\sim 90$\% of the monitored F-K stars with typical amplitudes $\le 0.1$\,mag. In all
these studies, the range of amplitudes is similar to our lightcurves, particularly after converting 
from V- to I-band. The detection rate, on the other hand, is 50-90\% and thus clearly higher than 
in our VLM sample.

The main observational selection effects that might influence the period detection rates are photometric accuracy,
sampling of the time series, and a biased choice of targets. The photometric accuracy in our campaigns
(particularly runs A and C) is better than in most aforementioned literature studies. In addition, most 
of the literature studies rely on fewer datapoints and are probably less sensitive to short periods.
Combining both effects, we would expect to find a larger fraction of periodic objects, not the opposite.

All our targets are selected from photometry alone and are thus unbiased with respect to the rotation/activity
properties. To our knowledge, the same applies to the samples observed in \citet{1996A&A...305..498A} and 
\citet{1999ApJ...516..263B}. The $\alpha$\,Per samples might be biased towards fast rotating
objects, since \citet{1993MNRAS.262..521O} selected presumed fast rotators for their monitoring study. 
Since the rotation-activity relation for periods of 1--5\,d is flat in these clusters 
\citep{1996ApJS..106..489P}, this does not imply a bias in activity levels. Overall, selection effects
do not appear to be responsible for the low number of stars with period in our sample. More likely,
this finding is caused by a mass dependence in the spot properties.

Monitoring studies in NGC2264 \citep{2003PhDT.........2L,2007prpl.conf..297H} and in the Pleiades 
\citep{2004A&A...421..259S} have pointed at a drop in photometric amplitudes by a factor of about 2 in the 
very low mass regime. This has been interpreted as evidence for a change in spot properties at 
$M\sim 0.3-0.5\,M_{\odot}$. We cannot confirm this drop in amplitudes in IC4665, which may be due 
to the small sample size. Indirectly, however, we confirm these findings, as low amplitudes could 
prevent period detection.

Possible explanations for low amplitudes and/or a low fraction of periodic lightcurves are a) very few spots, 
b) low contrast between spots and photosphere, c) many symmetrically distributed spots or one polar spot. 
From multi-filter monitoring \citet{2005A&A...438..675S} conclude that a) and/or c) are the most plausible
scenarios. We note that the current results from Zeeman Doppler imaging point towards large-scale, poloidal
fields for fully convective very low mass objects \citep{2008MNRAS.390..567M}. This indicates the presence 
of large polar spots and a lack of spots at low latitudes, which reinforces hypothesis c). 

\subsection{Long-term spot evolution}

With our three monitoring campaigns covering three years, we are for the first time able to probe
the long-term evolution of magnetic spots in VLM objects. Of particular relevance is the comparison
between the two WFI time series (run A and C): Having similar noise characteristics and sampling, they
allow us a direct check for long-term variations in the lightcurve.

The most important finding here is that out of 20 objects with detected period, only two show a
period in both campaigns. The lightcurves of those two objects will be further discussed in Sect. 
\ref{two}. In the majority of the cases, the characteristics of the spots have changed sufficiently 
after three years, preventing the confirmation of the period. Persistent spot activity is thus a rare 
phenomenon in young VLM objects. 

This result can again be compared with the properties of more massive stars. Large samples of 
solar-mass stars at ages 1--5\,Myr have been monitored repeatedly in the past, allowing us to put
constraints on the stability of the spot activity. \citet{2004AJ....127.1602C} have published
a five-year study of T Tauri stars in IC348. In agreement with other studies \citep[e.g.][]{2006AJ....132.1555N},
they find variations in amplitude and lightcurve shape between different seasons. In their
sample, 14 out of 27 stars with periodic variability exhibit the period in all five seasons;
23 of 27 stars in at least two out of five seasons. This is consistent with results from work in the Orion 
Nebula Cluster, where \citet{2001AJ....121.1676R} finds that about half of the periods measured in 1996/97 
can be recovered in a dataset obtained about a year earlier, although this earlier run had more 'patchy' 
sampling. Spot stability over several years has also been inferred from Doppler imaging studies 
\citep[e.g.][]{1999ApJS..121..547V}. 

Thus, for young solar-mass stars about half of the objects retain a detectable period over more than
one season. In contrast, only 10\% of the VLM objects in IC4665 exhibit a photometric period in both 
WFI campaigns separated by three years. This again points to a change in the spot properties in the
VLM regime, as discussed in Sect. \ref{spotmass}. In very low mass objects, spot activity causing 
photometric variability on the percent level is rare and transient. Since the amplitudes in our sample 
are small and in many cases close to the photometric noise limit, only a slight change in lightcurve amplitude, 
caused for example by a re-configuration of the spots, could lead to a non-detection in the period search. 
Thus, the lack of objects with stable periods may be a direct outcome of the drop in amplitudes in the
VLM regime.

\subsection{Objects with persistent periodic variability}
\label{two}

The two most interesting objects in our sample are probably candidates \#57 and \#59, both with persistent
periodic variability in all three observing campaigns, spanning three years. With masses $\sim 0.2\,M_{\odot}$
these two objects belong to the lowest mass stars known to date with long-term photometric variability. They
are thus prime targets for follow-up studies, e.g. multi-filter monitoring or Doppler imaging. 

\begin{figure*}
\includegraphics[width=4.0cm,angle=-90]{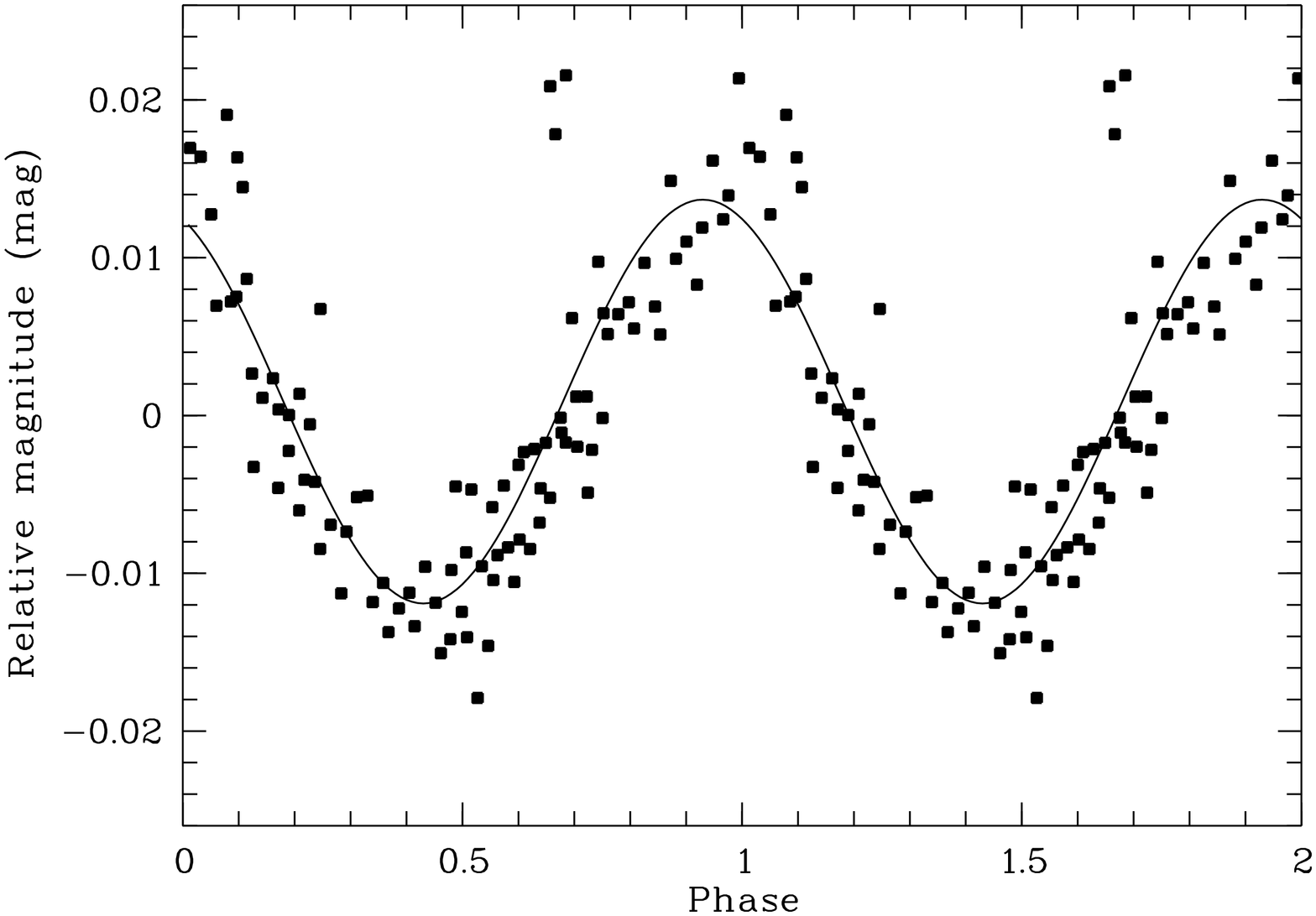} \hfill
\includegraphics[width=4.0cm,angle=-90]{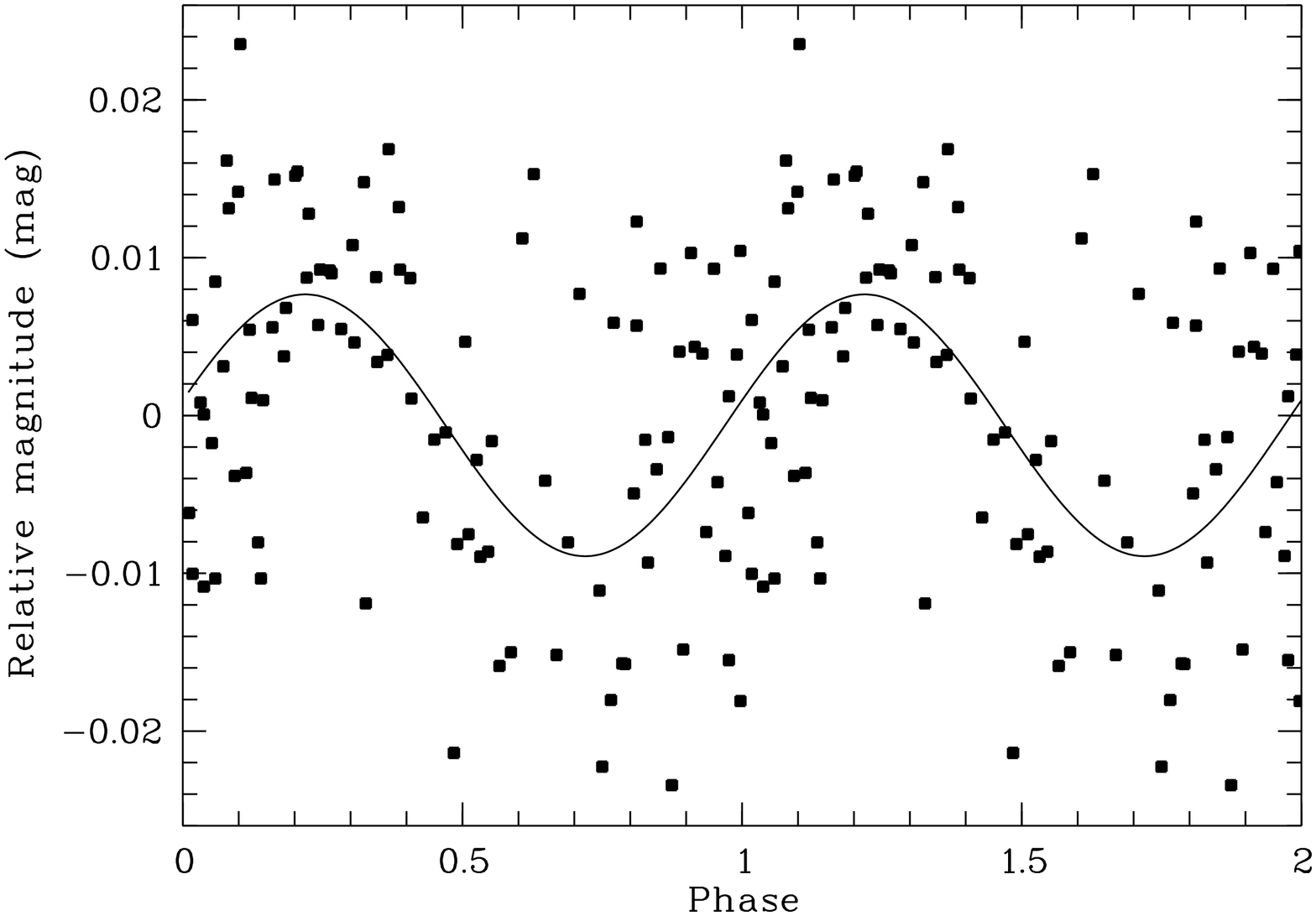} \hfill
\includegraphics[width=4.0cm,angle=-90]{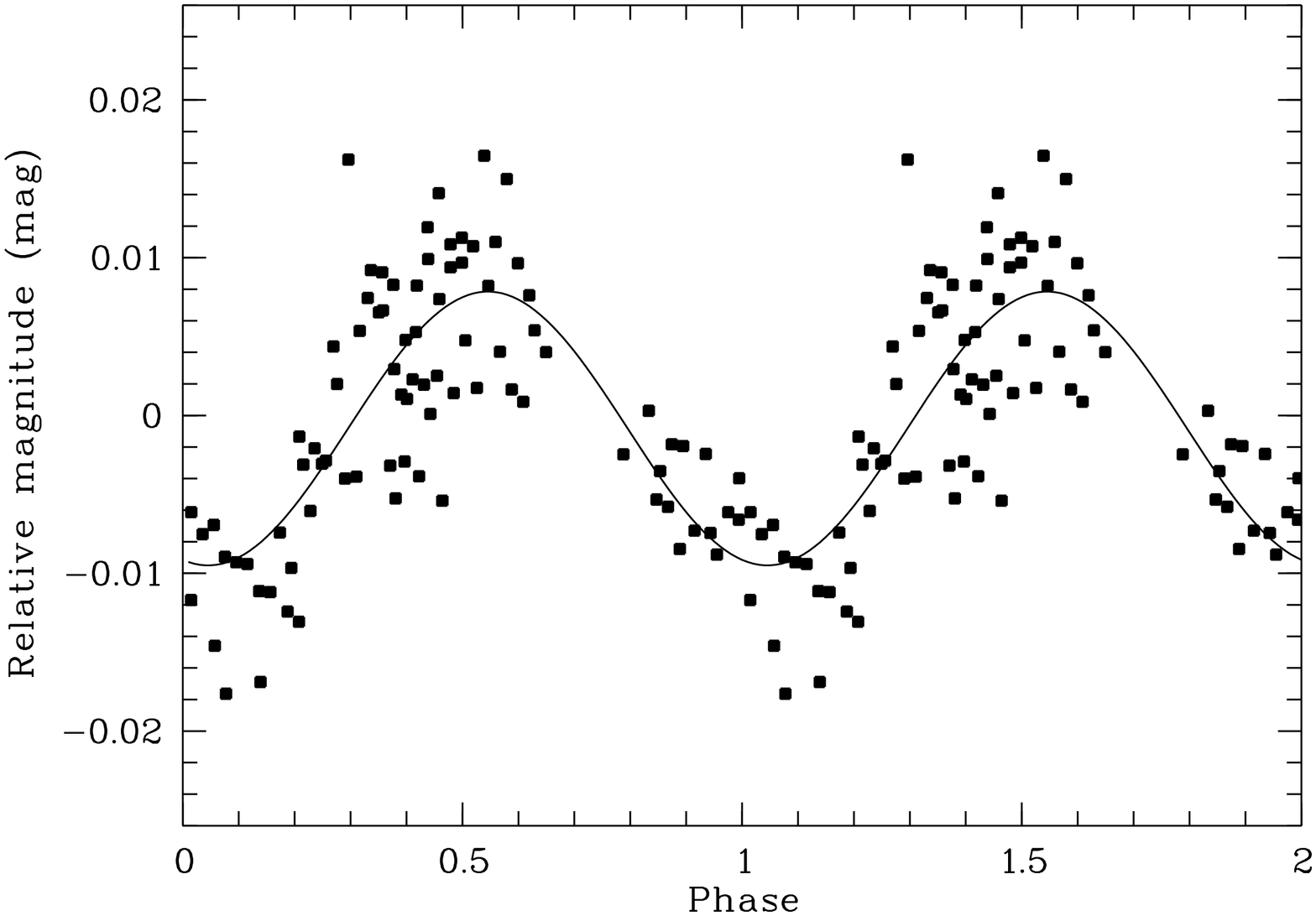} \\
\includegraphics[width=4.0cm,angle=-90]{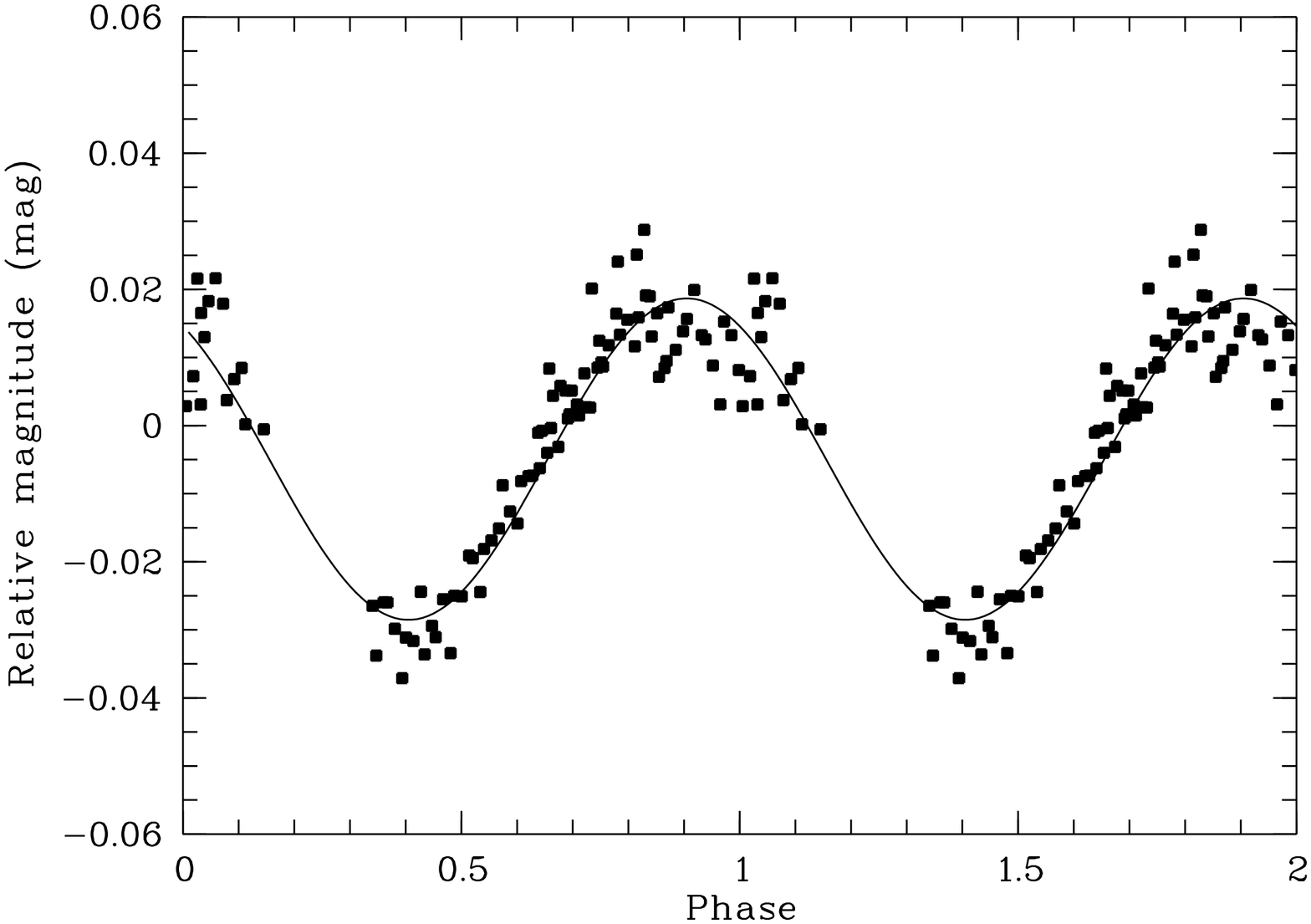} \hfill
\includegraphics[width=4.0cm,angle=-90]{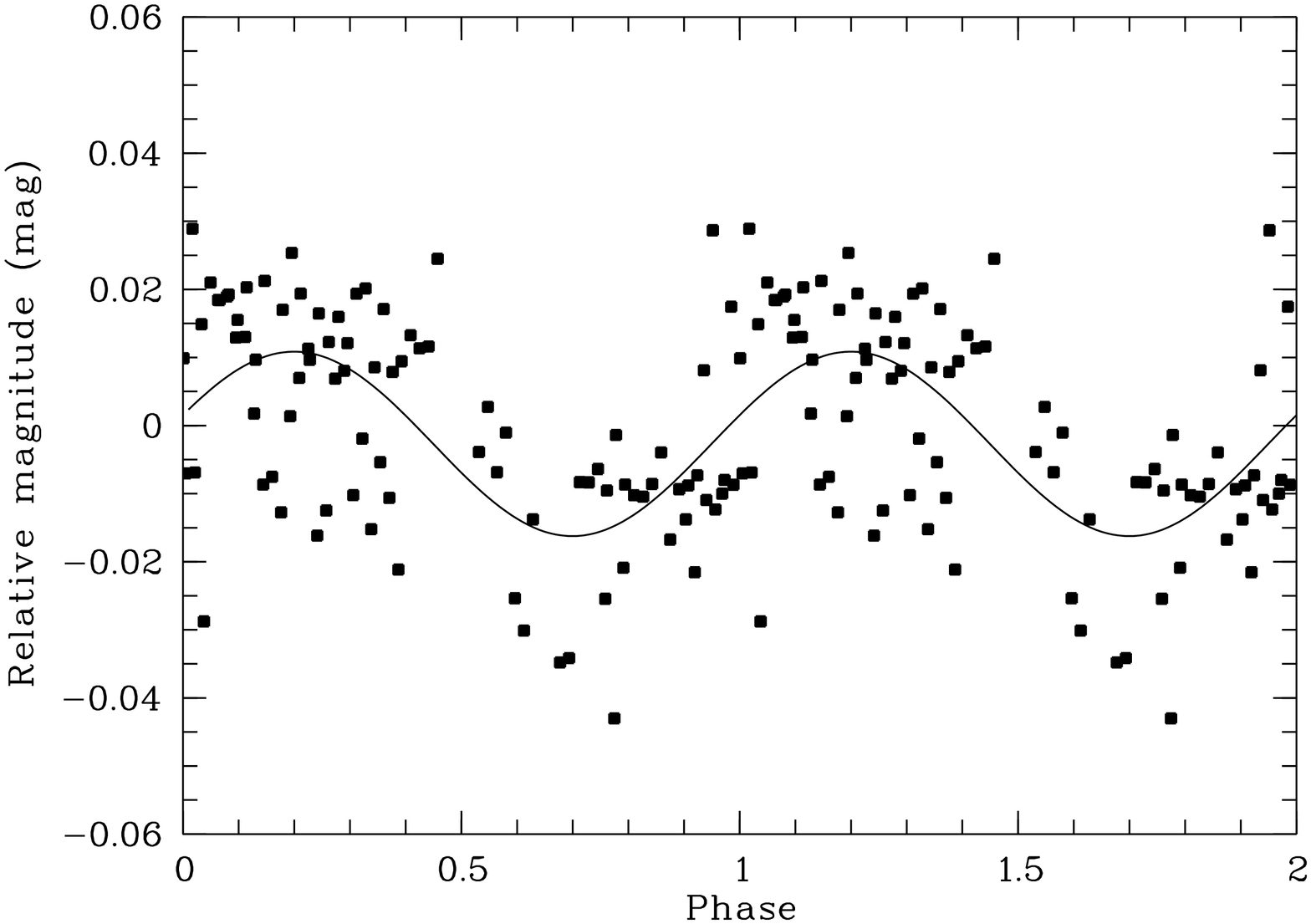} \hfill
\includegraphics[width=4.0cm,angle=-90]{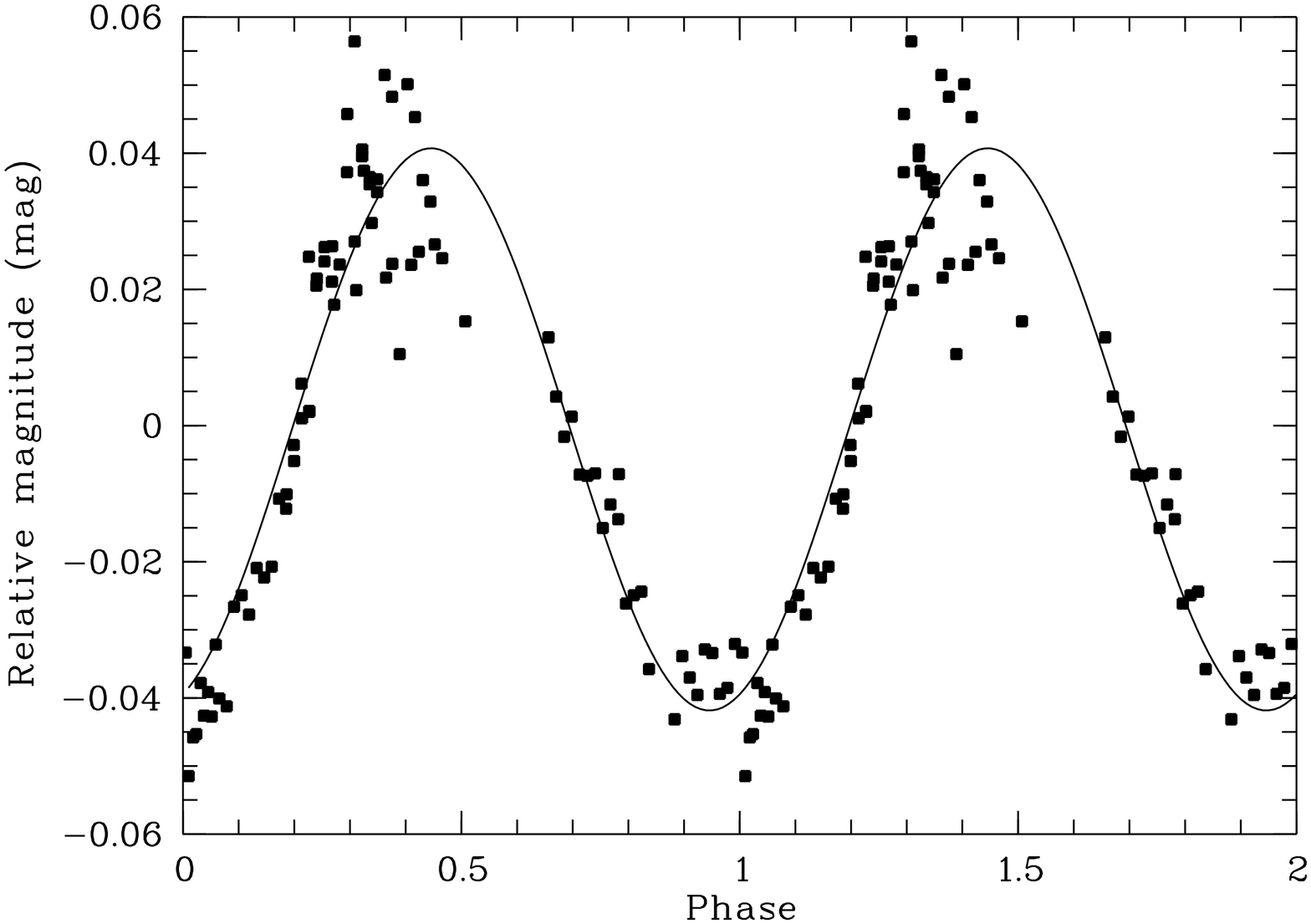} \\
\caption{Phaseplots for objects \#57 (upper panels) and \#59 (lower panels) for runs A to C (from
left to right). The lightcurves are plotted in phase to the assumed rotation period of $\sim 10$\,h
and $\sim 14.2$\,h, respectively. The best-fit sine function is overplotted, see Table \ref{perfit} 
for the fit parameters. Note the changing y-axis scale in the 3rd panel in the lower row. \label{f7}}
\end{figure*}

We re-determined the periods for \#57 and \#59 by fitting sine functions to the lightcurves from each 
run. The free parameters in the fit are phase, amplitude, and period, but the period was restricted 
to a small range around the assumed rotation period of 10 and 14\,h, respectively (see Sect. \ref{tls}). 
The fit results are summarised in Table \ref{perfit}. We do not attempt to fit all three seasons with a 
single periodic function, because the period uncertainty accumulated over a year or more causes a phase 
uncertainty larger than the period itself. In Fig. \ref{f7} we show for both objects the three lightcurves 
phased to the period given in Table \ref{perfit}. As pointed out earlier, the rotation periods are not 
recovered in run B, but for object \#57 the beat period at 18.6\,h is detected. The period fit for run
B are thus not considered trustworthy. 

\begin{table}
    \caption[]{Results from the period fitting to the lightcurves for objects \#57 and \#59. The period
    error $\Delta$P (h) has been determined following \citet{1986ApJ...302..757H}. The period fit did not
    yield significant results for the run B lightcurves; the periods from that run are considered
    unreliable.}
       \label{perfit} 
       \begin{tabular}{ccccc}
	  \hline
          ID & run  & P (h) & $\Delta$P (h) & A (mag)\\
          \hline
          57 & A    & 9.99  & 0.04 & 0.026 \\
	     & B    & 10.45 & --   & 0.017 \\
	     & C    & 9.69  & 0.04 & 0.017 \\
	  \hline
	  59 & A    & 14.04 & 0.04 & 0.047 \\
             & B    & 13.17 & --   & 0.027 \\
             & C    & 14.45 & 0.04 & 0.083 \\
	  \hline 
       \end{tabular}
\end{table}

In the following, we will discuss change in amplitude, lightcurve shape, and period over the three
seasons. For object \#57 the photometric amplitude is constant within the errorbars over all three seasons.
For object \#59, however, the amplitude shows striking variations: It drops by a factor of about two from
1999 to 2001 and increases then by a factor of about three between 2001 and 2002. The two most likely scenarios 
for such changes are a re-configuration of the magnetic spots and a strongly variable spot filling factor. 

\begin{figure}
\includegraphics[width=6cm,angle=-90]{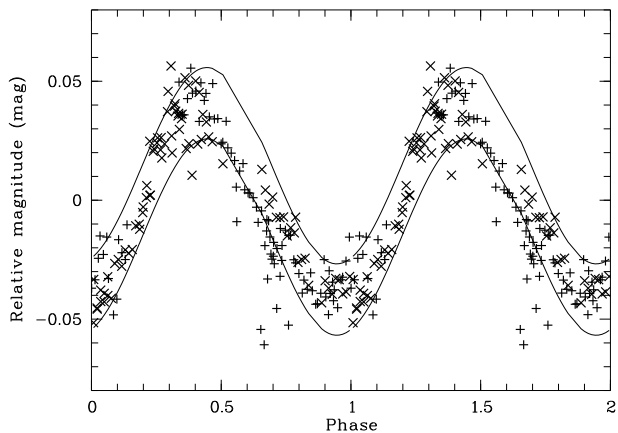} 
\caption{Combined datapoints from run A (plusses) and C (crosses) for object \#59. Both lightcurves are 
plotted in phase to the period measured in run C (14.45\,h). The lightcurve from run A has
been scaled by a factor of 1.77 to match the amplitude in run C. The two solid lines give the
range expected from the 1$\sigma$ photometric noise. \label{f8}}
\end{figure}

For both objects, the shape of the lightcurves are in good agreement with the sine function. This is not 
particularly surprising since the periods have been determined by trying to match a sine curve. The more 
relevant result is that the shape of the lightcurve does not change significantly over the three seasons.
For object \#59 this is demonstrated in Fig. \ref{f8}: We plot the datapoints from run A and run C in phase
to the period measured for run C (14.45\,h) scale the lightcurve from run A by a factor of 1.766 to match 
the amplitude of run C, and overplot them with the lightcurve measured in run C.

As can be seen in this plot, both sets of datapoints cover the same regions in phase space. The deviations
between both lightcurves are fully consistent with the expected scatter of about $\pm 0.015$\,mag, given as 
solid lines in Fig. \ref{f8}. This noise estimate includes the combined photometric noise of both campaigns
($\sim 0.01$\,mag), uncertainty in the period, and possible other sources of variabiliy (e.g., microflares).
Thus, within the uncertainties the lightcurves from run A and C are identical, except for a clear change in
amplitude. Since any re-configuration of the magnetic spots is expected to alter the shape of the lightcurve,
this favours a variable spot filling factor as explanation for the change in amplitude, maintaining the magnetic
configuration constant. Assuming that spot temperatures are on average constant, changes in the spot coverage 
by a factor of 2-3 on timescales of years are required to match the observations.

Magnetic cycles, as seen in solar-type stars, are one possible reason for a long-term variation in the spot 
filling factor. As of today, simulations of the magnetic field generation do not find strong evidence for 
the presence of cycles in fast rotating fully convective objects 
\citep{2006A&A...446.1027C,2006ApJ...638..336D,2008ApJ...676.1262B}. Our result might pose a challenge to these
models and certainly motivates continued monitoring of variable VLM stars.

In principle, our long-term study also allows us to probe differential rotation: If the 
rotation rate is a function of latitude, we expect to measure small changes in the rotation periods 
in different seasons, depending on the latitude of the dominant spot group. The rate of differential 
rotation can then be estimated simply from the scatter of the measured rotation rates $\Delta \omega$. 
Our candidates \#57 and \#59 show period changes of 0.3 and 0.4\,h between run A and C. This translates 
to $\Delta \omega$ of 0.4\,rad\,d$^{-1}$ for object \#57 and 0.3\,rad\,d$^{-1}$ for object \#59, 
corresponding to 2-3\% of the rotational velocity.

The formal errors from the period fit are smaller than that (0.04\,h), which would indicate a detection 
of significant differential rotation in those two objects. A more detailed look on the lightcurves, 
however, shows that the actual uncertainties in the period measurement are likely to be larger: Due 
to the lack of datapoints outside two consecutive nights, the periodogram peaks for run A are broad 
enough to allow for period uncertainties of $\sim 0.5$\,h. More specifically, the periods obtained for
run C give a convincing (albeit not the best) fit to the lightcurves from run A. Therefore, we consider
the periods from run A and C to be consistent for both objects, and treat the period differences as
upper limits for the rate of differential rotation.

From theoretical ground, fully convective objects are expected to show only little, if any, differential 
rotation \citep{2006A&A...446.1027C,2006ApJ...638..336D,2008ApJ...676.1262B}. Monitoring campaigns on 
T Tauri stars have confirmed this prediction \citep{2004AJ....127.1602C,2006PASP..118..828H},
with period changes rarely exceeding a few percent. For VLM objects there is growing evidence that 
the differential rotation rate is negligible, as found by Doppler imaging \citep[$<0.05$\,rad\,d$^{-1}$,][]{2005MNRAS.357L...1B}
and Zeeman Doppler Imaging \citep[$\sim$mrad\,d$^{-1}$,][]{2008MNRAS.384...77M,2008MNRAS.390..567M}. 
Our findings in IC4665 are certainly consistent with the expectations. Given the lack of objects 
with persistent variability in this mass regime, it is challenging to derive more stringent limits 
on differential rotation from photometric monitoring alone.

\section{Rotation periods in IC4665}
\label{rotation}

\begin{figure}
\includegraphics[width=6cm,angle=-90]{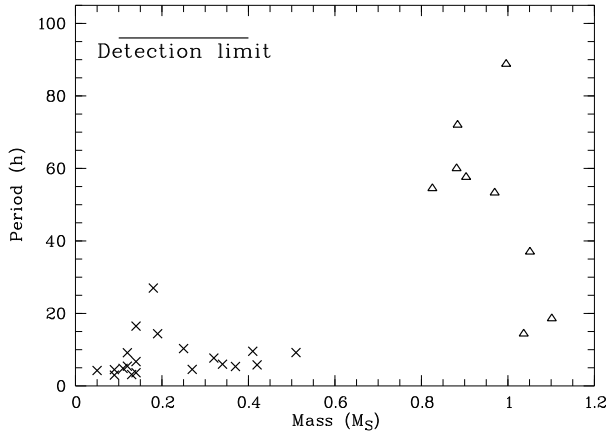} 
\caption{The total sample of rotation periods in IC4665, plotted vs. object mass. Crosses are periods
from this paper for VLM objects, triangles show periods from F-K stars \citep{1996A&A...305..498A}. The 
horizontal line in the top left corner indicates the approximate detection limit in our period search. 
All mass estimates have been derived in a consistent way and are thus comparable. \label{f9}}
\end{figure}

As established in Sect. \ref{spots}, the observed photometric periods correspond to the rotation
periods of our targets.\footnote{In rare cases, a particular spot distribution might generate a 
photometric period which is half the rotation period, e.g. when two similarly large features are 
located at opposite latitudes.} The sample of 20 new rotation periods for likely VLM members in 
IC4665 increases the number of objects with known periods in this cluster by a factor of three.
In Fig. \ref{f9} we plot the total sample of published periods in IC4665 as a function of mass, 
including our own periods and the results from \citet{1996A&A...305..498A} for F-K stars in this 
cluster. We note that the \citet{1996A&A...305..498A} variability campaign is likely not highly 
sensitive to very short periods ($<10$\,h), due to a small number of datapoints per observing night. 

The main feature in Fig. \ref{f9} is the cumulation of our datapoints at periods $<1$\,d, or conversely
the distinct lack of slow rotators in the VLM regime. Our period search may be incomplete, but it is 
definitely sensitive for periods up to 4\,d, as discussed in Sect. \ref{perser}. Thus, the observed
lack of slow rotators in the VLM regime is not induced by our sampling. This finding is supported by period 
samples in other young clusters -- VLM objects are almost exclusively seen as fast rotators, with the 
exception of objects with ages $<5$\,Myr, when disk-star interactions might still play a role. The
data in IC4665 provide another piece of evidence for a strong rotation-mass 
dependence.

To date we have no evidence that there is a population of slowly rotating VLM objects not found in 
the period searches. For the age and mass range considered here, photometric amplitudes are observed to 
be more or less independent of rotation, without any drop-off at long periods \citep{2008MNRAS.383.1588I},
which could cause us to miss long periods. We note that \citet{2009MNRAS.392.1456I} do in fact report 
such a drop-off for stars with $0.5<M<1.1\,M_{\odot}$, age 130\,Myr, and $P>10$\,d. We do not attribute 
too much relevance to this result in the context of the current study, for three reasons: a) The mass range 
and age are not comparable with our targets. b) Our study is focused on the period range $<4$\,d, for which
no drop-off in the amplitudes is reported. c) There is independent evidence for an upper period limit
well below 10\,d in the VLM regime.

This evidence comes from the available $v\sin i$ data obtained by high-resolution spectroscopy, which is not 
biased with respect to the spot activity. Albeit based on a limited sample, the lower envelope of $v\sin i$
in the Pleiades (120\,Myr) for $M<0.4\,M_{\odot}$ is well defined \citep{2000AJ....119.1303T}. In 
\citet{2004A&A...421..259S} we have shown that it translates into an upper limit for the rotation periods
of 1-2\,d, in good agreement with the available periods in the Pleiades as well as in IC4665.
Thus, the observed lack of slow rotators likely is a genuine effect and affects {\it all} VLM objects, 
and not only the ones with detected period. 

In the following subsections we discuss rotation in the VLM regime. In Sect. \ref{rotage} we focus on 
our new periods and evaluate the pre-main sequence rotational evolution based on the currently available data.
We find clear evidence for rotational braking on timescales of 5-50\,Myr. In Sect. \ref{bigpic}, 
we review the currently used wind braking laws for stars and brown dwarfs, before applying them to
pre-main sequence data in Sect. \ref{pms}. Finally, we put these results in context and identify problems
and future directions for research on rotation and magnetic properties (Sect. \ref{probs}).

\subsection{Tracing the pre-main sequence evolution}
\label{rotage}

\begin{figure}
\includegraphics[width=6cm,angle=-90]{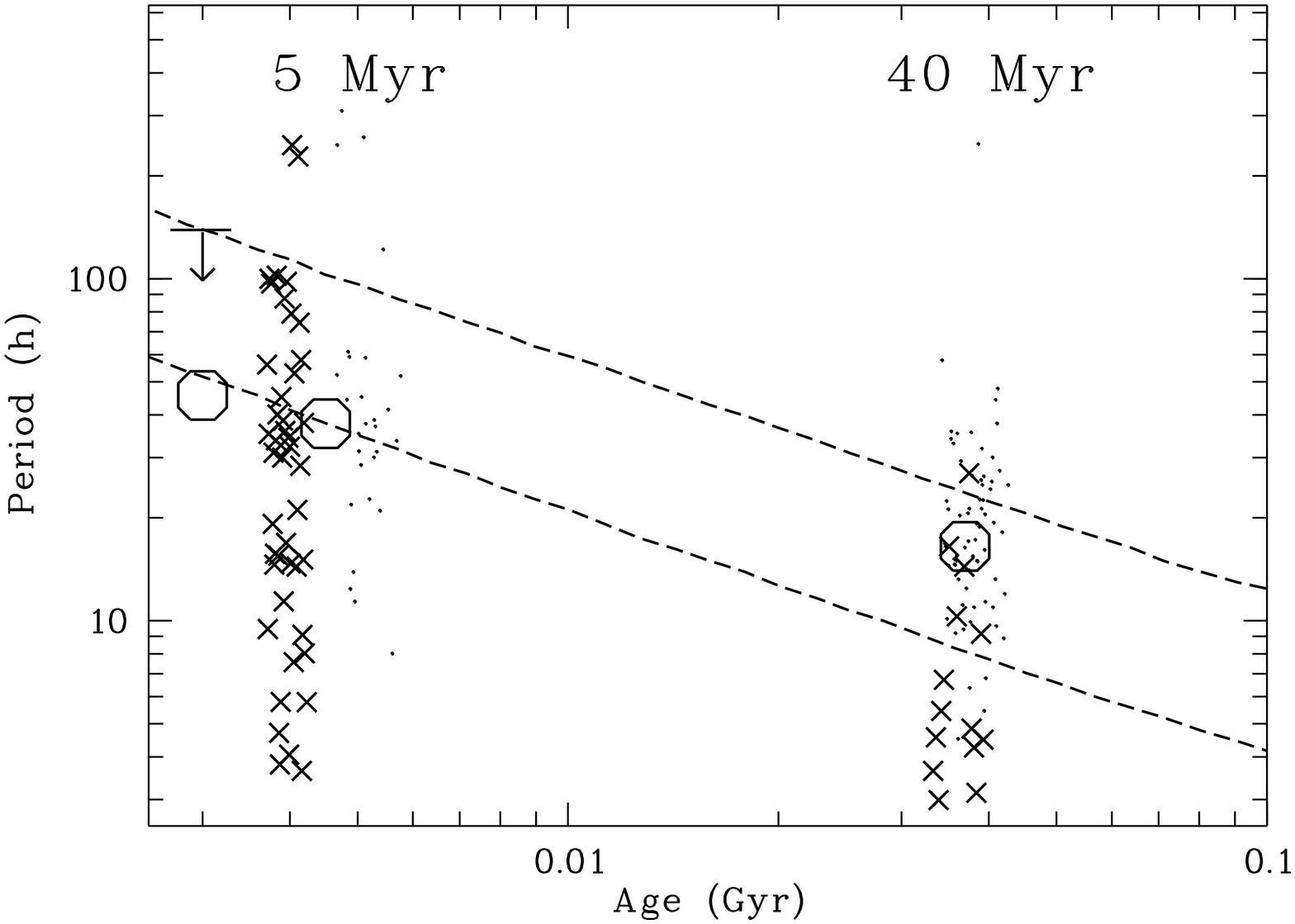} 
\caption{Rotational evolution of VLM objects in the pre-main sequence phase. Plotted are periods for
objects with $M<0.3\,M_{\odot}$ in the clusters $\sigma$\,Ori and $\epsilon$\,Ori 
\citep[average age $\sim 4$\,Myr,][]{2004A&A...419..249S,2005A&A...429.1007S}, NGC2362 
\citep[age $\sim 5$\,Myr,][]{2008MNRAS.384..675I}, IC4665 (this work), and NGC2547 
\citep[age $\sim 38$\,Myr,][]{2008MNRAS.383.1588I}. The period data for any given
cluster is plotted at a range of ages scattered randomly around the most likely cluster age,
for clarity. The large polygons show the median period for $0.1<M<0.3\,M_{\odot}$. In addition 
to the groups of clusters at 4--5 and $\sim 40$\,Myr, we plot median and 90\%-ile for VLM objects in 
NGC2264 at $\sim 3$\,Myr \citep[][$M<0.25\,M_{\odot}$ on their mass scale]{2005A&A...430.1005L}. The 
dashed lines show the period evolution assuming angular momentum conservation based on the evolutionary 
tracks by \citet{1998A&A...337..403B} for 0.3\,$M_{\odot}$ (upper track) and 0.2$\,M_{\odot}$ 
(lower track). \label{f10}}
\end{figure}

There is now a substantial sample of rotation periods for VLM objects in young open clusters.
With at least 20 VLM periods in seven pre-main sequence clusters (ONC, NGC2264, $\sigma$\,Ori,
$\epsilon$\,Ori, NGC2362, IC4665, NGC2547) we are in a position to put firm limits on the 
rotational braking at very young ages \citep[see][]{2007prpl.conf..297H,2008MNRAS.383.1588I}.
In Fig. \ref{f10} we plot periods for objects with $M<0.3\,M_{\odot}$ as function of cluster age. As
crosses we show periods in the clusters $\sigma$\,Ori \citep{2004A&A...419..249S}, $\epsilon$\,Ori
\citep{2005A&A...429.1007S} -- both plotted at an approximate age of $\sim 4$\,Myr --, and IC4665
(this paper, age $\sim 36$\,Myr). Periods from the {\it Monitor} project are plotted as small dots,
for the clusters NGC2362 \citep[age $\sim 5$\,Myr,][]{2008MNRAS.384..675I} and NGC2547 
\citep[age $\sim 38$\,Myr,][]{2008MNRAS.383.1588I}. In total, we base the following discussion on 
72 periods at $4-5$\,Myr and 82 periods at $\sim 40$\,Myr.

For the youngest clusters in this plot, the fraction of objects with disks and/or
accretion is $<10$\%, thus rotational braking due to disk/star interaction is not relevant here.
In all five clusters, the object masses have been determined in a consistent way, thus the cut-off
at $0.3\,M_{\odot}$ is consistently defined for all ages. For completeness, we also plot the median
and the 90\%-ile of the VLM periods in NGC2264 at age of 3\,Myr \citep{2005A&A...430.1005L}, noting 
that both masses and ages in this cluster are probably not fully comparable with the remaining 
datasets.

The monitoring campaigns in Orion and IC4665 are deeper than the {\it Monitor} data, extending into
the substellar regime. Combined with the well-established period-mass relation in the VLM regime,
this explains the fact that the period samples in these clusters extend to shorter periods. The upper
period limits, however, are defined by the highest mass objects in those samples and thus comparable.
To probe the evolution for a typical VLM star, we plot the median period for objects with 
$0.1<M<0.3\,M_{\odot}$ as large polygons, a mass regime where all variability campaigns are 
sensitive. For this purpose we join the samples at similar ages, i.e. Orion and NGC2362 (median 
37.7\,h, based on 57 objects) as well as IC4665 and NGC2547 (median 16.5\,h, based on 78 objects). 

The two dashed lines in Fig. \ref{f10} provide the comparison with the theory. They show the calculated 
rotational evolution assuming angular momentum conservation ($P(t) = P_0 \times (R(t)/R_0)^2$). 
Both tracks are based on the radii from the \citet{1998A&A...337..403B} evolutionary tracks for 
solar-metallicity stars. The upper track is for a 0.3$\,M_{\odot}$ star and can thus be compared with 
the upper period limit, while the lower track is for a 0.2$\,M_{\odot}$ star, meant to provide a 
comparison with the median values. 

Comparing periods and model tracks clearly shows that the observations are not in agreement with angular 
momentum conservation at timescales of 5-50\,Myr. The expected median period for a VLM star at $\sim 40$\,Myr 
is about 8\,h, while the observed median is approximately twice as high. Similarly, the upper period limit 
at 40\,Myr is not well-represented by the model: The track starts at $P_0\sim 100$\,h, with only 6 
objects ($8\pm 3$\%) have periods exceeding this value at young ages. At $\sim 40$\,Myr, however, 24 
out of 82 objects ($29\pm 5$\%) exceed the predicted period of $\sim 22$\,h. Excluding the highest 8\% 
of the periods at 40\,Myr yields an upper limit of about 36\,h, a factor of 1.6 too high compared with
the prediction for angular momentum conservation.

In summary, rotational braking clearly occurs for objects with masses of 0.1-0.3$\,M_{\odot}$ on timescales of 
$\sim 50$\,Myr. We see an offset between prediction for angular momentum conservation and observed periods 
by a factor of 1.6--2, implying a reduction in angular momentum by 40-50\% compared with the case without 
any braking. Note that this effect can hardly be explained by uncertainties in the radii from the 
\citet{1998A&A...337..403B} models: We only use the ratio of radii at different ages, thus any systematic 
problem with the models is likely to be eliminated. Moreover, radii from these models are found to be in 
good agreement with observations for early/mid M dwarfs \citep{2004ApJ...609..885M,2007ApJ...662.1254S}. 

\subsection{Wind braking in stars and brown dwarfs}
\label{bigpic}

Stellar winds powered by magnetic fieldsa are the main physical process governing the rotational braking on long 
timescales. Angular momentum losses due to stellar winds are commonly modeled based on the 
parameterisation by \citet{1988ApJ...333..236K}, which is derived from the analytical wind braking
law by \citet{1984LNP...193...49M}:
\begin{equation}
\frac{dJ}{dt} = -K\omega ^{1+4aN/3} \left(\frac{R}{R_{\odot}}\right)^{2-N}
\left(\frac{M}{M_{\odot}}\right)^{-\frac{N}{3}} \left(\frac{\dot{M}}{10^{-14}}\right)^{1-\frac{2N}{3}}
\end{equation}
Here $N$ is the 'wind index' describing the topology of the magnetic wind and $a$ is 
the dependence of the magnetic field on the rotation rate, 
$B\propto \omega^a$ \citep{1996ApJ...462..746B}. 

In most studies, a value of $N=1.5$ is chosen, corresponding to a mixture between dipole 
($N\sim 0.5$) and radial ($N=2.0$) field. By adopting $a=1.0$ and thus a linear relation between
B-field and rotation rate, one obtains $dJ/dt \propto \omega^3$ and reproduces the Skumanich law
$\omega \propto t^{-1/2}$, the empirical relationship for the decay of rotation rates of G-type 
stars on the main-sequence \citep{1972ApJ...171..565S}. 

This particular choice of the parameter $a$, however, fails to be applicable to the 'ultra-fast 
rotators' (UFRs) in young open clusters. In addition, it does not take into account that magnetic 
activity indicators are observed to saturate at high rotation rates 
\citep[see][and references therein]{1996ApJS..106..489P}. Therefore, it is usually assumed that 
the magnetic field saturates at a critical rotation rate, i.e. $a=0.0$ for $\omega >\omega_{\mathrm{crit}}$. 
With the same wind index as in the Skumanich case, Equ. 1 then yields 
$dJ/dt \propto \omega_{\mathrm{crit}}^2 \omega$ and $J(t) \propto \exp{(t/\tau)}$, with 
$\tau = 1/\omega_{\mathrm{crit}}^2$. By scaling $\omega_{\mathrm{crit}}$ appropriately with mass 
or effective temperature, models using this bimodal parameterisation of the angular 
momentum losses are indeed able to reproduce the main features in the available rotation period 
\citep[e.g.][]{1997A&A...326.1023B,2000ApJ...534..335s,2006MNRAS.370..954I} and $v\sin i$ data 
\citep[e.g.][]{2000AJ....119.1303T,2008ApJ...684.1390R}. 

Introducing a scaling of the critical rotation rate with mass or convective turnover timescale 
is justified, because the magnetic field generation is unlikely to be comparable in stars with 
different interior structure. Main-sequence stars with spectral types F to early M possess a 
radiative core and a convective envelope. The convection zone becomes progressively deeper with 
decreasing mass, so that all VLM objects are fully convective throughout their evolution. These 
changes are likely to have consequences for the dynamo processes generating the stellar magnetic 
fields. For example, it has been suggested that fully convective objects host an $\alpha^2$ dynamo 
\citep{2006A&A...446.1027C} or a distributed dynamo \citep{1993SoPh..145..207D} instead of the 
solar-type $\alpha\omega$ dynamo. \citet{2003ApJ...586..464B} proposed that the saturated ($a=0.0$) 
and linear regimes ($a=1.0$) of the angular momentum loss law represent objects with 'convective' and 
'interface' magnetic fields, called objects on the C-sequence or I-sequence. For consistency reasons, 
we will use this nomenclature in the following. 

\subsection{Application to pre-main sequence objects}
\label{pms}

Pre-main sequence VLM stars with ages between 10 and 100\,Myr are almost exclusively fast rotators and are 
thus considered to be on the C-sequence with $\omega >\omega_{\mathrm{crit}}$ \citep{2003A&A...397..147P}. 
The e-folding timescale for their wind braking is probably $>1\,$Gyr \citep{1998A&A...331..581D,2007MNRAS.381.1638S};
the angular momentum losses on the pre-main sequence depend primarily on radius with 
$\omega \propto R^{3/2}$, according to Equ. 1. This gives an angular momentum loss of $\sim 30$\% on 
timescales of 50\,Myr. As evaluated in Sect. \ref{rotage}, observations indicate losses of 40-50\%. 
Taking into account uncertainties in cluster ages and spindown timescale as well as the fact that a 
few objects might experience additional braking by star-disk interaction, this agrees well with the 
prediction.\footnote{Note that the dependence on radius in the wind law has been neglected in various 
previous publications, e.g. \citet{2004A&A...421..259S,2005ApJ...633..967H}.}

On the other hand, the observed estimates of angular momentum losses in the VLM regime
are nowhere near the effect expected for the I-sequence. Skumanich braking gives angular 
momentum losses of $>70$\% , i.e. significantly larger than the values seen in the VLM 
regime. This simply confirms that VLM stars on the pre-main sequence have to be considered
objects on the C-sequence. In summary, the rotational data for pre-main sequence VLM objects,
including our new periods in IC4665, fit the current paradigm for the rotation evolution,
as described in Sect. \ref{bigpic}.

In \citet{2007ApJ...662.1254S}, $v\sin{i}$ for F to early M stars in young stellar 
associations with ages between 6 and 30\,Myr have been analysed, a dataset that is 
complementary to the periods shown in Fig. \ref{f10}. Similarly to the VLM objects, the 
rotational pre-main sequence evolution of solar-mass stars is incompatible with Skumanich 
braking and allows only for relatively little angular momentum loss. Thus, in the pre-main 
sequence phase all stars seem to go through a C-sequence phase of rapid rotation and weak 
wind braking. They make the transition to the I-sequence and thus 'switch on' the Skumanich
braking at an age $t_{C-I}$, which is a function of mass. This has also been suggested in a 
recent paper by \citet{2009ApJ...695..679M}. According to the current data, $t_{C-I}$ is 
$50-100$\,Myr for 1.0$\,M_{\odot}$, 0.5-1\,Gyr for $M=0.5\,M_{\odot}$ \citep{2007MNRAS.381.1638S}, 
and several Gyrs for $M<0.3\,M_{\odot}$ \citep{1998A&A...331..581D}. The transition between 
I- and C-sequence generates a large gap in rotation rates between solar-mass and VLM stars 
on the main sequence.

\section{Two regimes of magnetic properties}
\label{probs}

\begin{figure}
\includegraphics[width=6cm,angle=-90]{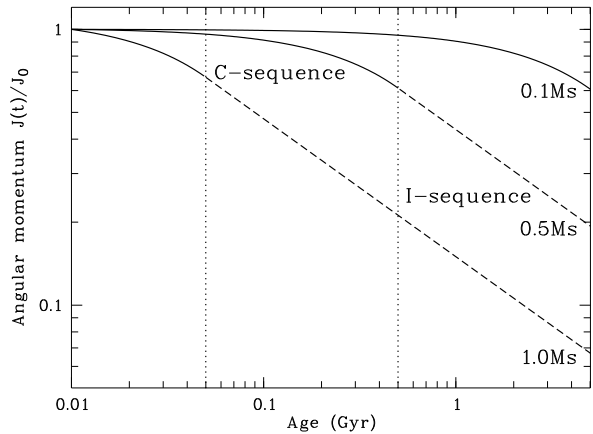} 
\caption{Schematic representation for the rotational evolution of stars at three different
masses. Solid and dashed lines show the spindown on the C-sequence and I-sequence, respectively.
The vertical dotted lines illustrate the transition from C- to I-sequence for 1$\,M_{\odot}$ (at
0.05\,Gyr) and 0.5$\,M_{\odot}$ (at 0.5\,Gyr). The effect of the pre-main sequence
contraction has been neglected here (see text). \label{f11}}
\end{figure}

In Fig. \ref{f11} we summarise the rotational evolution scheme described in the previous
Sect. \ref{rotation}. This figure is meant to serve as a qualitative illustration and does not
provide an accurate quantitative description. All stars start on the C-sequence with moderate wind 
braking in the pre-main sequence phase. At a given age $t_{C-I}$ which is a function of mass 
they make the transition to the I-sequence with Skumanich type spindown. In Fig. \ref{f11}, we 
plot the C-sequence in solid lines and the I-sequence in dashed lines. For $t_{C-I}$ we assume
0.05\,Gyr for 1$\,M_{\odot}$ and 0.5\,Gyr for 0.5$\,M_{\odot}$, consistent with the currently
available constraints (see Sect. \ref{pms}). Additionally, we assume that the e-folding timescale 
$\tau$ on the C-sequence scales with object mass. We neglect here the dependence of the 
rotational braking on the radius, which is relevant in the pre-main sequence phase.

With the I- and the C-sequence two regimes of wind braking have been identified. 
In order to explore the underlying physics of these regimes, we look for differences in the 
magnetic properties between C- and I-sequence. The age sequence defined by the transition from 
C- to I-sequence is similarly seen in the fall-off of chromospheric H$\alpha$ emission, which occurs 
at $\sim 50$\,Myr for G-K stars, at 0.5-1.0\,Gyr for early M stars \citep{1999ASPC..158...63H},
and at several Gyrs for M3-M5 stars \citep{2000ASPC..212..252H}. This may indicate that the onset 
of H$\alpha$ activity and the transition from C- to I-sequence are related, i.e. the 'switching 
on' of the Skumanich braking goes along with a drop in H$\alpha$ activity.

As already discussed in Sect. \ref{bigpic}, the rotation/activity relation is flat for very young 
stars at 1\,Myr \citep{2004AJ....127.3537S} and VLM objects \citep{2003ApJ...583..451M}, both C-sequence 
objects, and linear for slowly rotating solar-mass stars on the I-sequence \citep[see also][]{2003ApJ...586L.145B}. 
There are clear indications for a change of the magnetic field topology at the transition to fully convective objects 
\citep{2006Sci...311..633D}. Additionally, the characteristics of the photometric lightcurve, 
caused by magnetically induced spots, change significantly in the VLM regime (see Sect. \ref{spots}). 
Thus, C- and I-sequence are likely to constitute two different magnetic regimes. We summarise the 
currently available evidence for the bimodality in magnetic/rotational properties in Table \ref{ic}.

\begin{table}
    \caption[]{Qualitative summary of observable properties for objects on the C- and I-sequence. 
    For more references see text.}
       \label{ic} 
       \begin{tabular}{lcc}
	  \hline
           & C-seq  & I-seq \\
          \hline
          rotation                   & fast		       & slow		    \\
	  MS wind braking            & $\propto \exp{(t)}$$^a$ & $\propto \sqrt{t}$ \\
	  H$\alpha$ emission         & yes$^b$  	       & no/weak	    \\
	  rotation/activity relation & flat$^c$ 	       & linear 	    \\
	  differential rotation      & no/weak$^d$	       & strong 	    \\
	  magnetic topology          & poloidal$^e$	       & poloidal/toroidal  \\
	  spot activity              & weak$^f$ 	       & strong 	    \\
	  \hline
       \end{tabular}
       
       $^a$ \citet[][and references therein]{2003ApJ...586..464B}
       
       $^b$ e.g., \citet{1999ASPC..158...63H,2000ASPC..212..252H}
       
       $^c$ e.g., \citet{1996ApJS..106..489P}
       
       $^d$ \citet{2005MNRAS.357L...1B,2008MNRAS.384...77M}
       
       $^e$ \citet{2006Sci...311..633D,2008MNRAS.384...77M}
       
       $^f$ this paper
\end{table}

In the parametrisation for the wind braking given in Sect. \ref{bigpic}, two parameters
determine if an object belongs to the C- or I-sequence: rotation rate and mass. Since
both parameters are interdependent, one of them may be of more fundamental relevance than
the other. It is particularly tempting to explain the whole body of rotation/magnetic 
properties based exclusively on stellar mass, because it determines the interior structure, as 
pointed out in Sect. \ref{bigpic}: While VLM objects and very young stars are fully convective, 
main-sequence F-K stars harbour a radiative core. This difference may be the fundamental reason 
for the two regimes of rotational/magnetic properties. 

However, as already discussed in \citet{2007MNRAS.381.1638S}, the mass and age limits for 
the transition from C- to I-sequence do not coincide with the prediction for the occurrence 
of the radiative core; thus there is no clear-cut connection between the change in internal 
structure and the change in rotation/magnetic properties. For all objects that do not remain 
fully convective throughout their evolution (i.e. $M>0.35\,M_{\odot}$), the radiative core in 
F-K stars is thought to appear at $<$10-30\,Myr and is fully developed after $<$100-200\,Myr 
\citep{1997A&A...327.1039C,1994ApJS...90..467D}, significantly earlier than the transition
from C- to I-sequence. To maintain the notion that internal structure is essential for 
magnetic properties, one has to assume that following the formation of a radiative core, 
it takes a considerable timespan before the characteristics of the I-sequence 
(see Table \ref{ic}) become observable.

If interior structure is the underlying reason for the two regimes, VLM objects --
fully convective throughout their evolution -- are not supposed to leave the C-sequence. This is 
clearly at odds with the observations. Several studies have identified mid/late M dwarfs with 
very low or undetected $v\sin{i}$ and activity levels below the range expected for the saturated 
regime \citep{2000ASPC..212..252H,2003ApJ...583..451M}. Albeit fully convective, these VLM stars 
are likely to constitute the non-saturated regime, suggesting that slowly rotating VLM objects 
do exist and can operate an I-type (Skumanich) rotational braking. From these simple arguments it 
seems unlikely that interior structure (or mass) alone can account for the two different regimes 
of magnetic/rotational properties. Thus, two parameters, rotation and mass, are required to 
distinguish between I- and C-sequence.

The fact that we observe two regimes of magnetic properties points to a more general problem in the 
current models, as described in Sect. \ref{bigpic}. The braking laws for I- and C-sequence both 
assume $N=1.5$, a wind structure somewhere between dipole and radial field. By holding this 
parameter constant, it is implicitly assumed that all low-mass stars and brown dwarfs share 
the same type of magnetic field structure, with saturation on the C-sequence being the only 
difference. As outlined above, there is clear evidence for a change of magnetic field properties 
at the transition from C- to I-sequence (see Table \ref{ic}), which makes the choice of constant 
$N$ questionable. The specific benefit of the value $N=1.5$ is that is conveniently cancels out the
mass loss rate in the wind braking law, which is difficult to constrain by observations. On the other
hand, the explanatory power of an angular momentum loss law that does not include the mass loss rate
is probably limited.

For a more satisfying understanding of the underlying physics, it may be needed to re-investigate 
the parametrisation of the wind braking law in Equ. 1. For solar-mass stars, \citet{1996ApJ...462..746B} 
originally proposed two ways to explain the ultrafast rotators in young open clusters, objects that 
we would classify as being on the C-sequence: either with saturation ($a=0.0$ instead of 1.0) or 
with a low value for the wind index until 100\,Myr ($N<1$, close to a dipole field) and then switching 
to the Skumanich type $N=1.5$. The second type of parametrisation has not been explored in detail 
and may turn out to be closer to the physical reality. While saturation of magnetic activity at 
high rotation rates is a well-established fact \citep[see][and references therein]{1996ApJS..106..489P}, 
it is probably the filling factor that saturates \citep{1996IAUS..176..237S}. When choosing $a=0.0$ in 
the wind braking law, however, we assume that the magnetic field strength is saturated. It may 
be necessary to resort to models with $a$ and $N$ as free parameters, thus accomodating both 
for saturation and a change in the magnetic wind structure.

The choice of $N=1.5$ may not be appropriate for the Skumanich type I-sequence either: Recent numerical
simulations by \citet{2008ApJ...678.1109M} indicate a significantly smaller value for this parameter 
(about 0.7 instead of 1.5). This would imply that angular momentum losses depend at least weakly on 
the mass loss rate ($dJ/dt \propto \dot{M}^{0.5}$), which is probably a more plausible description than
one without dependence on $\dot{M}$. 

From these considerations is becomes clear that the currently available theoretical framework for the stellar
spindown is lacking a satisfactory description of the physics of the rotation-wind connection. Future
work should aim to verify the bimodal wind braking law as given in Equ. (1) and explore possible alternatives.
Observationally, the emerging problems encourage to obtain more reliable constraints on mass loss rates, as
this may turn out to be crucial to improve the modeling of rotational evolution. Having said this, the 
bimodal schematic used in current models and described in detail above has been proven to have considerable 
predictive power. Given a proper calibration, it allows us to use rotation rates as indirect indiator of 
stellar ages and masses.

\section{Conclusions}

We have monitored a large sample of very low mass objects and brown dwarfs in the young cluster IC4665 over three 
seasons, aiming to measure rotation periods and to assess the long-term evolution of spot activity. For 20
objects photometric periods have been derived, which likely represent the rotation periods. Two objects show 
persistent variability. For these two, we find a consistent period that is detected in all three datasets.
However, this is not necessarily the most significant period in each dataset. This illustrates that periods 
for individual objects obtained from a single monitoring campaign might not always represent the rotation 
period. The periods range from 3 to 30\,h, with photometric amplitudes from a few mmag to 0.12\,mag. We
find that we are sensitive to periods longer than 30\,h, i.e. the lack of longer periods is not a bias
in our dataset. This indicates a lack of slow rotators among VLM objects in this cluster. We confirm that 
flares are very rare events in very low mass objects.

Compared with solar-mass stars, the very low mass objects in IC4665 show weak photometric variability with
low amplitudes, which could be due to a change of spot properties at the fully convective boundary, for
example to a more symmetric spot distribution. Persistent variability is rare, again in contrast to the
results for more massive stars. We do not see evidence for differential rotation. One object with persistent
periodic variability changes its amplitude by a factor of two over the course of 3 years, while the shape
of the lightcurve remains the same, which could be explained by a long-term magnetic cycle. Our sample of 
lightcurves in IC4665 provides a fundament for future studies of the long-term evolution of activity in the
VLM regime.

We find that our periods fit into a scenario where VLM objects experience moderate wind braking in the pre-main
sequence phase. They are inconsistent with the strong Skumanich-type wind braking. This is in line with the
current paradigm for the angular momentum evolution of low-mass stars, which includes a bimodal nature of the
wind braking. We point out that this bimodality is also seen in various properties associated with the magnetic
activity, indicating a more fundamental physical difference. While the current parameterised description of 
wind braking is successful in reproducing main features in the period distributions in open clusters, it does 
not provide a satisfactory physical picture. 

\section*{Acknowledgments}

We would like to thank Ansgar Reiners, Sean Matt, Jonathan Irwin, and Jerome Bouvier for instructive 
discussions regarding topics related to this paper. This work was partially supported by 
{\it Deutsche Forschungsgemeinschaft} (DFG) grant Ei\,409/11-1 and 11-2.
 
\newcommand\aj{AJ} %Astronomical Journal
\newcommand\araa{ARA\&A} %Annual Review of Astron and Astrophys
\newcommand\apj{ApJ} %Astrophysical Journal
\newcommand\apjl{ApJ} %Astrophysical Journal, Letters
\newcommand\apjs{ApJS} %Astrophysical Journal, Supplement
\newcommand\aap{A\&A} %Astronomy and Astrophysics
\newcommand\aapr{A\&A~Rev.} %Astronomy and Astrophysics Reviews
\newcommand\aaps{A\&AS} %Astronomy and Astrophysics, Supplement
\newcommand\mnras{MNRAS} %Monthly Notices of the RAS
\newcommand\pasa{PASA} %Publications of the Astron. Soc. of Australia
\newcommand\pasp{PASP} %Publications of the ASP
\newcommand\pasj{PASJ} %Publications of the ASJ
\newcommand\solphys{Sol.~Phys.} %Solar Physics
\newcommand\nat{Nature} %Nature
\newcommand\bain{Bulletin of the Astronomical Institutes of the Netherlands}

\bibliographystyle{mn2e}
\bibliography{aleksbib}

\label{lastpage}

\end{document}